\documentclass[a4paper,fleqn]{cas-dc}

\usepackage[numbers]{natbib}
\usepackage{graphicx}
\usepackage{enumitem}
\PassOptionsToPackage{table}{xcolor}
\usepackage[table]{xcolor}
\usepackage{ifthen}
\usepackage{xspace}
\usepackage{subfigure} 
\usepackage{booktabs} 
\usepackage{threeparttable}  
\usepackage{soul}

\definecolor{codegreen}{rgb}{0,0.6,0}
\definecolor{codegray}{rgb}{0.5,0.5,0.5}
\definecolor{codepurple}{rgb}{0.58,0,0.82}
\definecolor{backcolour}{rgb}{0.95,0.95,0.92}

\DeclareRobustCommand{\head}[1]{\noindent\textbf{#1.}}

\newcommand{\changed}[1]{\textcolor{black}{#1}}

\def\tsc#1{\csdef{#1}{\textsc{\lowercase{#1}}\xspace}}
\tsc{WGM}
\tsc{QE}

\begin{document}
\let\WriteBookmarks\relax
\def\floatpagepagefraction{1}
\def\textpagefraction{.001}

\shorttitle{A Multi-Year Grey Literature Review on AI-assisted Test Automation}

\shortauthors{Ricca et al.}  

\title [mode = title]{A Multi-Year Grey Literature Review on AI-assisted Test Automation}  

%

\author[1]{Filippo Ricca}[orcid=0000-0002-3928-5408]

\cormark[1]

\ead{filippo.ricca@unige.it}

\ead[url]{https://person.dibris.unige.it/ricca-filippo/}


\affiliation[1]{organization={University of Genoa},
            addressline={Via Balbi 5}, 
            city={Genova},
            postcode={16126}, 
            country={Italy}}
            
\author[2]{Alessandro Marchetto}[orcid=0000-0002-6833-896X]




\affiliation[2]{organization={University of Trento},
            addressline={Via Sommarive 9}, 
            city={Trento},
            postcode={38123}, 
            country={Italy}}

\cortext[1]{Corresponding author}

\author[label3,label4]{Andrea Stocco} 
[orcid=0000-0001-8956-3894]

\affiliation[label3]{organization={Technical University of Munich},
            addressline={Boltzmannstra{\ss}e 3}, 
            city={Munich},
            postcode={85748}, 
            country={Germany}}

\affiliation[label4]{organization={fortiss GmbH},
            addressline={Guerickestra{\ss}e 25}, 
            city={Munich},
            postcode={80805}, 
            country={Germany}}


\begin{abstract}
\textbf{}\textbf{Context:}
Test Automation (TA) techniques are crucial for quality assurance in software engineering but face limitations such as high test suite maintenance costs and the need for extensive programming skills. Artificial Intelligence (AI) offers new opportunities to address these issues through automation and improved practices. 

\noindent
\textbf{Objectives:}
\changed{Given the prevalent usage of AI in industry, sources of truth are held in grey literature as well as the minds of professionals, stakeholders, developers, and end-users.}
To this aim, our study surveys grey literature to explore how AI is adopted in TA, focusing on the problems it solves, its solutions, and the available tools.
Additionally, the study is complemented by expert insights.

\noindent
\textbf{Methods:}
Over five years, we reviewed over 3,600 grey literature sources, including blogs, white papers, and user manuals, and finally filtered 342 documents to develop taxonomies of TA problems and AI solutions. We also cataloged 100 AI-driven TA tools and interviewed five expert software testers to gain insights into AI's current and future role in TA.

\noindent
\textbf{Results:}
The study found that manual test code development and maintenance are the main challenges in TA. In contrast, automated test generation and self-healing test scripts are the most common AI solutions. We identified 100 AI-based TA tools, with Applitools, Testim, Functionize, AccelQ, and Mabl being the most adopted in practice. 

\noindent
\textbf{Conclusion:}
This paper offers a detailed overview of AI’s impact on TA through grey literature analysis and expert interviews. It presents new taxonomies of TA problems and AI solutions, provides a catalog of AI-driven tools, and relates solutions to problems and tools to solutions. Interview insights further revealed the state and future potential of AI in TA. Our findings support practitioners in selecting TA tools and guide future research directions.
\end{abstract}

\begin{keywords}
Test Automation \sep Artificial Intelligence \sep AI-assisted Test Automation \sep Grey Literature \sep Automated Test Generation \sep Self-Healing Test Scripts
\end{keywords}

\maketitle

\section{Introduction}\label{sec:introduction}

The development of Test Automation (TA) techniques~\cite{TA} is meant to advance the quality assurance (QA) processes for software engineers~\cite{2016-Leotta-Advances,neonate}. TA supports a wide range of testing tasks, including automated code analysis, unit testing, integration testing, acceptance testing, and performance testing, applicable to various software products such as web and mobile applications. 
However, the limitations of TA frameworks like, e.g., Selenium WebDriver~\cite{selenium} and Selenium IDE~\cite{neonate} have become evident when creating complex test suites. These tools still demand substantial testing knowledge and programming skills, providing limited assistance in producing high-quality test scripts. Activities like developing robust locators~\cite{2016-Leotta-JSEP} and deterministic test scripts are still predominantly manual processes~\cite{2016-Leotta-Advances}. Additionally, maintaining automated test scripts becomes laborious and challenging due to constantly changing requirements and software evolution, resulting in issues like flaky or fragile tests that make test scripts costly to maintain~\cite{6200127,2021-Ricca-SOFSEM}. 

The integration of Artificial Intelligence (AI) and Machine Learning (ML) into TA is getting significant attention from researchers and practitioners, recognizing AI's potential to bridge the gap between human and machine-assisted testing activities~\cite{AITA,RiccioEMSE20}. In this paper, we refer to these techniques as Artificial Intelligence assisted Test Automation (AIaTA). AI holds the promise of transforming TA by simplifying or automating various testing activities, including test planning, authoring, development, and maintenance. Despite the growing adoption of AIaTA by companies~\cite{2021-Ricca-NEXTA}, there remains a limited understanding of the challenges it addresses, the solutions it offers, and the existing tools and their integration with the software development process.

Several secondary studies have reviewed existing work on AIaTA~\cite{9514942,8351734,9141124,10.1145/3551349.3563240,enase20}. \changed{This paper is a secondary study that focuses on the grey literature~\cite{GAROUSI2019101} to capture practitioners' perspectives on the adoption of AIaTA. The main goal of our work is to understand the available AIaTA tools, the problems they address, and the innovative solutions they offer, from the developers' perspective.} 

\changed{In our previous work~\cite{Quatic2023,2021-Ricca-NEXTA}, of which this article is an extension, we surveyed the grey literature to collect, consolidate, and organize existing AI practices for TA. 
This paper builds upon the grey literature reviews by Ricca et al.~\cite{2021-Ricca-NEXTA,Quatic2023} by incorporating more recent sources, thus presenting an extended, multi-year grey literature review. We provide details on the differences between the prior papers and this article. 
First, this article contains an expanded empirical evaluation (\autoref{sec:empirical study}) as we extended the study to sources from the year 2024. This comprehensive review spans several years, demonstrating the generalizability of the results and examining the underlying relationships among TA problems, AI-based solutions, and existing tools.
Second, to gather further insights on the application of AIaTA, we conducted interviews with five researchers and practitioners to collect their experience. These interviews were recorded, transcribed, and analyzed to identify the key problems, solutions, and tools discussed by the interviewees. 
Thus, the main contribution of this extended paper is a set of taxonomies of problems, solutions, and tools regarding AIaTA, which are corroborated by experts. To our knowledge, this is the first work that includes interviews with developers on the usage of AI for TA.}

Our experience shows that grey literature is a valuable yet underutilized resource for AI practices in TA, containing insights that practitioners may not have the time or scientific expertise to rigorously extract. Our goal is to surface these valuable insights, often found in scattered documentation or within the knowledge of professionals, stakeholders, developers, and end-users. We believe our work can help practitioners understand the current state and practices in AIaTA, aiding in the selection of appropriate tools for their testing needs. Additionally, our findings can guide researchers in identifying issues that need further investigation and new research directions. 

This paper is organized as follows: \autoref{sec:background} provides essential background information to help comprehend the rest of the paper. This includes an overview of the current limitations of TA and an introduction to an AI-based mechanism, self-healing test scripts, which addresses the well-known issue of fragile tests. \autoref{sec:empirical study} reports research questions, adopted procedure, document selection phase, and data analysis of our multi-year grey literature review. 
The results are detailed in \autoref{sec:results}, and \autoref{sec:discussion} explores these findings in relation to the threats to validity and empirical evidence collected. \autoref{sec:interviews}, on the other hand, briefly presents the design and results of the interviews conducted with five expert software testers. Finally, \autoref{sec:relatedworks} summarizes related works, and \autoref{sec:conclusions} concludes the paper.
\section{Background}\label{sec:background}

In this section, background information is provided to understand the content of the paper. Specifically, a brief explanation is given of what test automation is, along with the known practical benefits and drawbacks as described in both practice and literature. Subsequently, the rationale behind the utilization of AI and ML in this context is explained, along with the associated benefits. Finally, to provide a more concrete understanding, the process of self-healing test scripts is detailed. This process utilizes ML to automatically adjust failing test scripts due to the evolution of the application under test.

\subsection{Test Automation}

Test automation is the practice of using specialized software tools or frameworks to control the execution of tests and compare actual outcomes with predicted outcomes~\cite{7888399}. It encompasses the entire testing process within an organization, aiming to improve efficiency, accuracy, and coverage in software testing~\cite{10.5555/2132831}. The primary component driving test automation is the use of test scripts---i.e., programs that run specific portions of the software being tested. These test scripts perform a sequence of predefined actions against the Application Under Test (AUT), consisting of commands and inputs. The expected results of test scripts are typically described by assertions, which are specific statements provided by a testing framework, e.g., JUnit~\cite{10.5555/961868}. Assertions check values, e.g., the result of the call under test, or the final status of some part of the system, against given conditions and raise an exception in case of failure. The testing framework detects these exceptions and marks the tests as failed. 

Automated test scripts can target various levels of software~\cite{10.5555/2161638}, addressing different aspects of the system to ensure comprehensive testing coverage. At the most granular level, unit tests focus on individual components or functions within the AUT, verifying that each unit behaves as expected in isolation. This level of testing is crucial for identifying and fixing bugs early in the development process.

Moving up the hierarchy, automated tests can target APIs (Application Programming Interfaces). API testing ensures that the interactions between different software components, as well as with external services, are functioning correctly~\cite{10.1145/3533767.3534401}. This involves sending requests to the API endpoints and validating the responses, including data formats, status codes, and content. API testing is essential for verifying the integrity and reliability of the communication paths within the AUT.

At the highest level, automated test scripts can perform system-wide testing in an end-to-end, user-focused manner. End-to-end (E2E) testing simulates real user scenarios and interactions with the software, covering the entire application flow from start to finish~\cite{2016-Leotta-Advances}. This approach ensures that all components and subsystems work together seamlessly to deliver the intended user experience. E2E tests are particularly prevalent in web and mobile environments, where user interactions span multiple layers of the application, including the user, back-end services, and databases.
This approach is especially important in these kind of environments due to the complexity and variability of these platforms. For example, web applications must be tested across different browsers~\cite{10.1145/1985793.1985870} and devices to ensure consistent performance and user experience. Similarly, mobile applications must be tested on various operating systems, screen sizes, and hardware configurations. Automated end-to-end tests help identify issues that may arise from these variations, ensuring that the application remains working and user-friendly across different environments.

By targeting different levels of the software, from individual units to entire systems, automated test scripts provide a comprehensive and scalable testing solution. This multi-level approach~\cite{8402699} helps ensure that all aspects of the software are thoroughly tested, contributing to the overall quality and reliability of the final product.

Test automation offers numerous benefits~\cite{Rafi2012BenefitsAL}, including increased testing efficiency, higher test coverage, and improved accuracy by reducing human error. It enables repetitive and regression testing~\cite{630875}, where tests are run frequently to ensure that new changes do not negatively impact existing functionality. This is crucial for maintaining software quality over time.

However, test automation also presents well-known challenges, in particular in the context of web and mobile applications~\cite{NASS2021106625,ICST2023}. Traditional automation testing faces issues such as slow test execution and the persistent problem of maintaining test scripts. Slow test execution is a primary reason for delays in testing, often caused by a focus on GUI automation, poorly designed test scripts, and insufficient test case sequencing. 

Excessive test maintenance is another major issue, as test scripts are highly sensitive to the application's UI and structure. Any minor change in the UI necessitates corresponding changes in the test script, leading to a significant portion of automation effort being dedicated to test script maintenance. 

Additionally, the test script can break due to small changes in the code (test script fragility problem), such as renaming or relocating a GUI component on the screen~\cite{2018-Stocco-FSE}. Managing and maintaining test data adds another layer of complexity, requiring testers to create test data generation scripts and use version control effectively. Finally, traditional automation testing is complex and code-intensive, and a lack of skilled resources often leads to the failure of test automation projects due to inadequate planning.

\subsection{AI-assisted Test Automation}

Artificial Intelligence involves enabling computer programs to execute tasks that would typically necessitate human intelligence.
Within this broad definition, we find machine learning (ML), which involves pattern recognition and learning from data to solve classification or regression problems~\cite{10.5555/1162264}. Additionally, Computer Vision (CV) provides techniques for analyzing and understanding images, much like how humans perceive them, with popular applications including pattern recognition, image analysis, and optical character recognition. Finally, Natural Language Processing (NLP) enables computers to analyze and understand human language. In the last ten years, the availability of large benchmarks of labeled data and unprecedented computing power has enabled the application of Deep Learning techniques such as deep neural networks (e.g., convolutional neural networks and recurrent neural networks) to complex vision and spatiotemporal problems. More recently, transformer architectures have enabled the exploitation of language embedding that is at the basis of large language models (LLMs), i.e., novel types of AI that achieved unprecedented performance in NLP and coding tasks.

Traditionally, AI methods and models have been widely applied to various phases of the software development lifecycle, including software testing. The application of AI to support software testing is a well-established and increasingly popular research topic, as evidenced by several recent studies in the literature~\cite{10.1145/3616372,8638573,8717439,10.1007/978-3-031-27499-2_18}.

In the context of Test Automation, AI is also employed for various purposes~\cite{10.1145/3551349.3563240,enase20}. For example, AI and ML algorithms are used to analyze application behavior and user interactions to automatically generate test scripts. This reduces the manual effort required for test script creation and improves test coverage. Another example is self-healing test scripts, in which test constructs are automatically maintained when the test code becomes obsolete or loses sync with the AUT during software evolution. This is meant to diminish the maintenance overhead associated with test automation. Additionally, AI techniques can generate additional test data by analyzing patterns observed in production data. This facilitates the creation of diverse test scenarios, for more comprehensive testing of the AUT. Lastly, ML algorithms can detect anomalies in test results, including unexpected behavior or performance issues. This allows for early detection of defects, mitigating their impact on the production environment. Additionally, AI can enhance test execution by intelligently scheduling tests, considering factors such as code changes, risk profiles, and resource availability. 

\section{Empirical Study}\label{sec:empirical study}

Our research centers on the grey literature that discusses the application of AIaTA. We consider the following  research questions:

\noindent
\head{RQ\textsubscript{1} (Issues/problems)}
\textit{How does AI contribute to mitigating challenges in TA?}

\noindent
\head{RQ\textsubscript{2} (Solutions/approaches)}
\textit{In what ways does AI provide solutions to enhance TA?}

\noindent
\head{RQ\textsubscript{3} (Tools and Platforms)}
\textit{Which AI tools/platforms are widely used in the context of software testing?} 

\noindent
\head{RQ\textsubscript{4} (Problems vs. Solutions)}
\textit{What is the relationship between TA problems and solutions in AIaTA?}

\noindent
\head{RQ\textsubscript{5} (Tools vs. Solutions)}
\textit{How are AIaTA tools and solutions interconnected?}

RQ\textsubscript{1} identifies the main TA problems and issues that are supported by AI-based solutions, i.e., AI-enhanced testing techniques and tools. To answer RQ\textsubscript{1}, we manually built a taxonomy of TA issues and problems addressed by AI-based solutions, considering the sources of grey literature.

RQ\textsubscript{2}  analyzes the solutions proposed for TA problems and issues that are based on AI algorithms and tools. To answer RQ\textsubscript{2}, we built a taxonomy of AI-based solutions used to address TA issues and problems identified in RQ\textsubscript{1}. In this case, as well, the information was inferred using grey literature sources.

RQ\textsubscript{3} identifies the existing tools and platforms that assist in the testing phase with AI-based solutions presented in the literature.
To answer RQ\textsubscript{3},  we extracted from grey literature existing tools that implement the AI-based solutions identified in RQ\textsubscript{2}.

RQ\textsubscript{4} maps TA problems and issues with proposed AI-based solutions. To answer RQ\textsubscript{5},  we mapped the problems and issues identified in RQ\textsubscript{1} with their proposed AI-based solution identified in RQ\textsubscript{2}.

Finally, the last research question RQ\textsubscript{5} maps such proposed AI-based solutions towards existing tools that implement them. To answer RQ\textsubscript{6}, we mapped the proposed AI-based solution identified in RQ\textsubscript{2} with the implemented tools supporting them, as listed in RQ\textsubscript{3}.

\begin{figure*}[t]
\centering
\includegraphics[trim=0cm 10cm 0cm 0cm, clip=true, width=\linewidth]{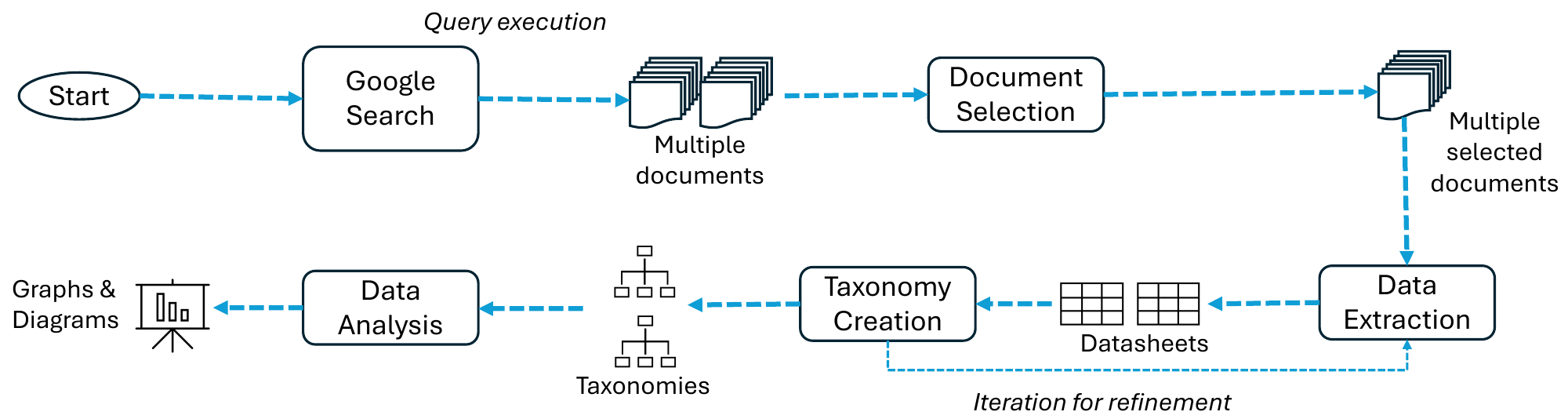}
\caption{Overview of the procedure adopted in each iteration of the grey literature.}
\label{fig:search}
\end{figure*}

\subsection{Procedure}\label{sec:procedure}

\changed{To review the grey literature relevant to our research questions, we carried out a selection procedure designed according to the guidelines by Garousi et al.~\cite{GAROUSI2019101}.} Particularly, we iterated the five-step procedure defined in our initial publication~\cite{2021-Ricca-NEXTA} three times, with an approximately one-year interval between each iteration, and further refined it in our extended study~\cite{Quatic2023}.
\autoref{fig:search} shows these five steps: 

\begin{enumerate}

\item Google search: by using a string query, we searched for potentially interesting documents.

\item Document selection: by applying inclusion/exclusion criteria, we checked the initial set of potentially interesting documents, thus identifying a reduced number of documents of interest, to be further analyzed.

\item Data extraction: we read and analyzed each document of interest to extract information concerning TA problems, AI-based solutions, and tools. 

\item Taxonomies creation: we created taxonomies of problems, solutions, and available tools.

\item Data analysis: we further analyzed the collected information using Sankey diagrams~\cite{Schmidt2008} and manual in-depth analysis for discovering relationships.  

\end{enumerate}

In the rest of the section, we present each step in detail.

\subsubsection{Google Search}

The Google search was intentionally broad to gather a large pool of documents. This approach was intended to maximize the retrieval of all relevant documents, at the cost of including documents that were not directly relevant to our study. 
Thus, we established specific inclusion and exclusion criteria to filter out the irrelevant documents and ensure that the remaining ones align with the study's scope. We used Google in incognito mode to conduct our searches, to ensure that our search results were not influenced by previous search histories or personalized algorithms.
For our Google search, we used the following query string:

\vspace{1mm}
\noindent\fbox{%
    \parbox{0.95\linewidth}{%
		\center{((``artificial intelligence'' OR ``AI'' OR ``machine learning'' OR ``ML'') \par AND (``test automation'' OR ``automated testing''))
    }}%
} 
\vspace{2mm}

\changed{The first part of the query string is characterized by words related to artificial intelligence and machine learning, whereas the second part is related to automated testing. All relevant documents contained an instance of each keyword from each part of the string (AND operator), whereas keywords within the same part were ORed. Hence, we have applied eight queries overall in each of the three repetitions of the search.}
\changed{For each query, the first 15 pages of results were scraped, each having 10 documents. No relevant documents were found after the 15$^{th}$ page. The eight applied queries led us to collect 1,200 documents for each repetition (150 documents for each query), thus resulting in a total of 3,600 documents collected and then analyzed in the three repetitions.}
The output of this step is a list of candidate web documents potentially regarding TA conducted with AI-based solutions. 
\changed{\autoref{tab:data1}, among other aspects, reports the number of documents collected in this step for each iteration (column ``\# Docs'') and the date on which the search was conducted (column ``\# Date'').}

\subsubsection{Document Selection}

The list of candidate web documents has been analyzed according to a set of inclusion and exclusion criteria, for filtering out irrelevant documents.
We mainly considered three inclusion criteria: (i)~the document needs to investigate AI or ML tools or methods that can support TA; (ii)~the document should apply to either capture-replay, programmable (or script-based), visual, or combinations of these testing approaches; (iii)~tools' websites and presentations are included as long as they specify useful information, and (iv)~presentations and slide decks are included only if in scope. We mainly excluded documents (exclusion criteria) if: (i)~they are scientific peer-reviewed papers; (ii)~they are not written in the English language; (iii)~they provide guidelines for using AI within manual testing; (iv)~they are videos or books; (v)~they are part of the website that requires registration; and (vi)~they contain only generic information on AI or software testing, without relating the two concepts. 
\changed{These criteria have been applied by analyzing each document obtained in the initial search with Google, thus only those documents that provide direct evidence about the study's objective were retained. The output of this step is the list of relevant selected documents regarding TA conducted with AI-based solutions.} 
\changed{\autoref{tab:data1} (column ``\# Selected'') reports the number of documents selected for the analysis in this step for each iteration, among the ones collected in the Google search step. In parentheses, we have reported the cumulative total value across the various iterations. We can see that, in each repetition, we selected around 10\% of the collected web documents, for an overall selection of about 9.5\% (i.e., 342 out of 3,600 documents) of documents identified in the Google Search phase.}

\begin{table}[t]

\caption{Summary of data related to the document search and taxonomies construction.}
\label{tab:data1}

\centering
\scriptsize

\setlength{\tabcolsep}{1pt}
\renewcommand{\arraystretch}{1}

\resizebox{\columnwidth}{!}{ %

\begin{tabular}{lccccccccccc}

\toprule

\textbf{Iteration} 
& \textbf{Date}  
& \textbf{\#Docs}  
& \textbf{\#Selected} 
& \textbf{\#Non-Con.} 
& \textbf{\#Problems} 
& \textbf{\#Solutions} 
& \textbf{\#Tools} \\

\midrule 

First~\cite{2021-Ricca-NEXTA} & Sept. 2020 & 1,200 & 156 & 14 & 35 & 35 & 50 \\
Second~\cite{Quatic2023} & Febr. 2023 & 1,200 & 95 (251) & 4 & 5 (40) & 9 (44) & 17 (67) \\
Third & Apr. 2024 & 1,200 & 91 (342) & 6 & 7 (47) & 15 (59) & 33 (100) \\

\bottomrule

\end{tabular}
}
\end{table}

\subsubsection{Data Extraction}

In this step, each relevant candidate document has been read and analyzed in detail and the following information has been collected:
(1)~web link to the document, (2)~name and surname, if any, about the author(s) of the document, (3)~the publication date of the document (if available), (4)~the type of the document (e.g., blog post, interview transcript), (5)~the test automation tool(s) investigated or described in the document, (6)~the testing level discussed (e.g., unit, integration, system, acceptance), (7)~the test automation problems and issues addressed, and (8)~the solution offered. 
\changed{\autoref{tab:data3} reports a summary of the most frequent types of documents among the selected ones, e.g., blog posts, and interview transcripts (column ``Document Type'' in the top part of the table). We can see that most of the documents were web articles or blogs. Overall, we were able to classify most of the document, i.e., 76.9\%.}
\changed{\autoref{tab:data3} reports also the number of documents in which the author's name was not specified (row ``Without author(s)'' in the bottom part of the table). We can see that it was impossible to identify the name of the authors for a non trivial number of documents, overall for 37.1\% of documents, with an increased trend in the last repetition.}

\changed{For each iteration, the authors conducted a pilot study by labeling a randomly selected sample of 10 documents. Each author labeled independently the sample of documents (e.g., type of TA problems and TA solutions), then during a meeting each document that was labeled differently by at least two authors was re-analyzed by all the authors together and the labels were discussed for reaching an agreement. These discussions were also used to refine the descriptions of the different categories of the taxonomies, thus reducing ambiguous definitions. Only in a few of the compared cases (4.5\%), we started the discussion for reaching an agreement since we observed different labels. Instead in the other cases (95.5\%), we observed the same labels for at least two classifiers, out of three, and for all three classifiers in the majority of the cases (59.1\%). Generally, hence, we observed that the consensus on the classification of the documents was high and the discussions quite limited, allowing the authors to proceed with the analysis independently on separate sets of documents for the subsequent phases. }

\begin{table}[t]

\caption{Summary of types of documents.}
\label{tab:data3}

\centering
\scriptsize

\setlength{\tabcolsep}{2pt}
\renewcommand{\arraystretch}{1}

\resizebox{\columnwidth}{!}{ %

\begin{tabular}{lccc}

\toprule

\textbf{Document Type} 
& \textbf{\# First~\cite{2021-Ricca-NEXTA}} 
& \textbf{\# Second~\cite{Quatic2023}} 
& \textbf{\# Third} \\

\midrule 
Blog									& 38	& 24	& 35  \\
Interview transcript	& 0		& 2		& 0   \\
Manual								& 0		& 2		& 0   \\
Post									& 2		& 7		& 8   \\
Slide									& 8		& 0		& 0   \\
StackOverflow					& 2		& 0		& 0   \\
Technical feed				& 0		& 0		& 3   \\
Tool									& 6		& 3		& 1   \\
Web article						& 46	& 40	& 30  \\
White-paper						& 5		& 0		& 0   \\ [0.5em]
\cline{2-4}
\multirow{2}{*}{Total}	&108	(69.2\%) &	78 (82.1\%)	&	77 (84.6\%) \\
													& \multicolumn{3}{c}{263 (76.9\%)} \\
\cline{1-4}
\multirow{2}{*}{Without author(s)}	&	 53 (33.9\%)		& 	22 (23.1\%)  &  52 (57.1\%)\\
																		& \multicolumn{3}{c}{127 (37.1\%)}  \\

\bottomrule

\end{tabular}
}
\end{table}

\begin{figure*}[t]
\centering
\includegraphics[width=\linewidth]{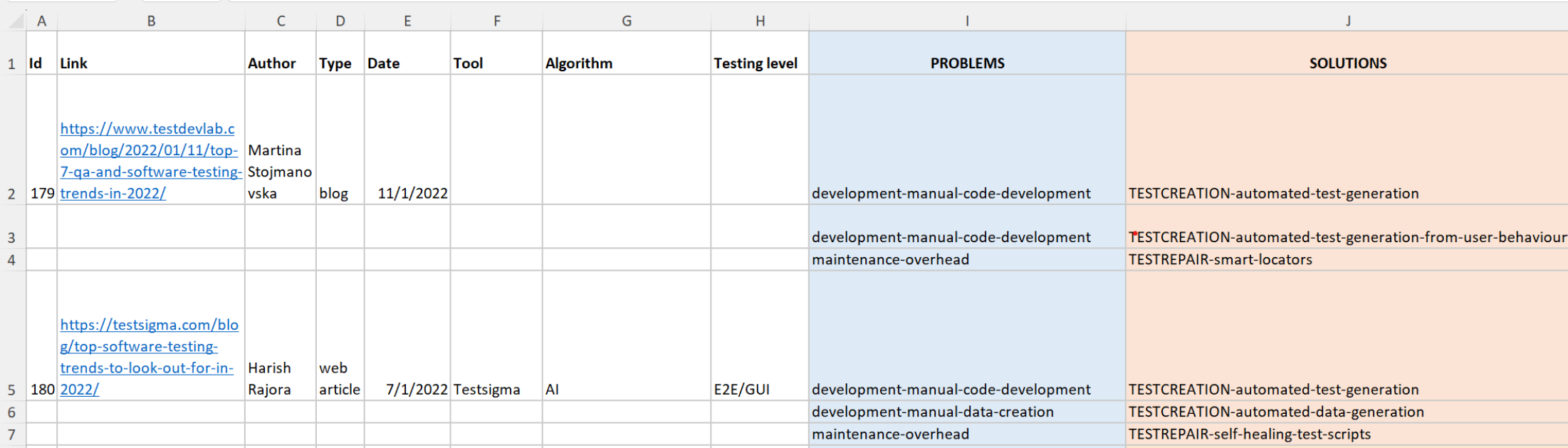}
\caption{A screenshot of the publicly available Google Docs spreadsheet featuring the data extracted from the documents.}
\label{fig:snapshot}
\end{figure*}

In detail, we build a tabular representation of the data extracted from the documents implemented as an online spreadsheet on Google Docs. 
\autoref{fig:snapshot} shows an image of the online spreadsheet on Google Docs we used to gather data, which details the contribution aspects.
Each row of the table reports the information collected for a document and an individual problem or solution. Hence, more rows can be used to detail several TA problem-solution tuples found in a single document. 
During the mapping process, the authors reused existing labels whenever applicable to avoid introducing nearly identical labels for the same TA problem/solution and to maintain consistent naming conventions.
We recall that the set of TA problems and TA solutions considered has been defined iteratively by starting from a small set of documents (as documented in~\cite{2021-Ricca-NEXTA}). Each newly identified problem and/or solution was incrementally considered, and the document was re-analyzed if needed.

\changed{In the three iterations, not all documents contributed to the identification of relevant TA problems, solutions, and tools supporting them. \autoref{tab:data1} (column ``\# Non-Con.'') reports the number of documents that did not contribute to the taxonomies in each iteration. We can notice that the reported numbers are rather limited and, in particular, they decrease between the first iteration and the subsequent ones, meaning that the selected documents tend to discuss more and more interesting content.}

\begin{table}[t]

\caption{Summary of Credibility and Content Quality.}
\label{tab:data4}

\centering
\scriptsize

\setlength{\tabcolsep}{2pt}
\renewcommand{\arraystretch}{1}

\resizebox{\columnwidth}{!}{ %

\begin{tabular}{lcccc}

\toprule

& \textbf{First~\cite{2021-Ricca-NEXTA}} 
& \textbf{Second~\cite{Quatic2023}} 
& \textbf{Third}
& \textbf{Total} \\

\midrule 

\bf Credibility \\ [0.1em]

\quad Low  		& 39 (42.8\%) & 39 (57.3\%) & 36 (42.8\%)  & 12 (4.5\%)\\
\quad Medium  & 33 (36.2\%) & 22 (32.2\%) & 33 (39.2\%)  & 104 (39.8\%)\\
\quad High  	& 19 (20.8\%) & 7 (10.2\%)  & 15 (17.8.7\%) & 145 (55.5\%)\\ [0.5em]

\bf Content Quality \\ [0.1em]

\quad Low  		& 8 (8.1\%) & 2 (4.1\%) & 1 (1.1\%) & 114 (46.9\%)\\
\quad Medium  & 32 (32.6\%) & 35 (47.9\%) & 37 (41.1\%)  & 88 (36.2\%)\\
\quad High  	& 58 (59.1\%) & 35 (47.9\%) & 52 (57.7\%) & 41 (16.8\%)\\

\bottomrule

\end{tabular}
}
\end{table}

\changed{It is worth noting that the quality of documents typically found in grey literature, such as those analyzed in this study, tends to be more diverse and often more challenging to evaluate compared to traditional systematic literature reviews. It is also more complex to conduct a sort of quality assessment of the used sources. By inspired the quality assessment checklist reported by~\cite{GAROUSI2019101}, we focused on: (i) the presence of authors; (ii) the quality and understandability of the content of the document; and (iii) a subjective assessment of the credibility of the document as perceived by the classifiers, i.e., the authors in our case. For each iteration, we collected this information for each selected document. In particular, the content of the document and the document's credibility are scored in terms of Low/Medium/High according to the judgment of the classifiers. We recall that \autoref{tab:data3} (row ``Without author(s)'' in the bottom part of the table) reports the number of documents for which an author was not identified.}

\changed{\autoref{tab:data4} summarizes the results in terms of content quality and document credibility as assessed by the authors. We can notice that overall only a limited amount of documents have been classified as Low in terms of credibility (i.e., 4.5\%). 
However, the classifiers did not provide any judgment for approximately 23\% of the documents (i.e., 342 selected documents minus 12+104+145), suggesting uncertainty or low credibility for these sources. Concerning the document content quality, we can notice that overall a large set of documents have been classified as of Low quality (i.e., 46.9\%). Moreover, the classifiers did not express any judgment for about 28\% (i.e., 342 selected documents minus 114+88+41) of the documents, for which we can assume a doubt to low content quality. These numbers suggest that, on one hand, the overall quality of the considered sources is sufficient for analysis; on the other hand, they confirm that grey literature is not always reliable~\cite{GAROUSI2019101}.}

\subsubsection{Taxonomy Creation}

By collecting the identified tuples of test automation problems and solutions, as well as the list of existing tools, two taxonomies have been constructed to answer RQ\textsubscript{1} and RQ\textsubscript{2}, following a systematic process~\cite{2020-Gyimesi-STVR}, while the list of tools has been used to answer RQ\textsubscript{3}.
To build the taxonomies, we clustered related TA problems and/or TA solutions, thus identifying categories of problems and solutions. Then, parent categories have been created, by following a specialization relationship between categories and subcategories. Especially, at the beginning of the process, some iterations between this step and the previous one, related to Data extraction, were needed to clearly identify the relevant information. 
The last columns of \autoref{tab:data1} report, for each iteration, the number of identified categories of problems (column ``\# Problems'') and solutions (``\# Solutions'') in the two taxonomies and the number of identified tools (``\# Tools''). In parentheses, we have reported the cumulative total value across the various iterations.

\subsubsection{Data Analysis}

We answered the first three research questions (RQ\textsubscript{1}, RQ\textsubscript{2} and RQ\textsubscript{3}) by analyzing the two taxonomies of problems and solutions and the list of tools.
Further analysis has been conducted after the construction of the taxonomies by considering the information related to the tuples of TA problems, solutions, and tools. In particular, several Sankey diagrams~\cite{Schmidt2008} have been constructed aiming at visualizing the possible interconnections between AI-based solutions and TA problems (to answer RQ\textsubscript{4}), and AI-based tools and AI-based solutions (to answer RQ\textsubscript{5}). 
In a Sankey diagram, the relations between nodes are shown by links that connect input and output nodes (in our case, respectively, solutions--problems and tools--solutions), while the width of a link indicates the relevance or magnitude of the relationship. By using the Sankey diagrams, we identified trends and patterns in the relations among TA solutions, problems, and tools, thus being able to answer RQ\textsubscript{4} and RQ\textsubscript{5}. 
More specifically, we employed \texttt{pandas} data-frames~\cite{mckinney2010data} to filter information related to the problems addressed and solutions provided (RQ\textsubscript{4}), as well as the tools used (RQ\textsubscript{5}). We excluded entries where the problem, solution, or tool was either unspecified or too generic. For the relevant entries, we counted each pairwise combination of \texttt{<problem, solution>} and \texttt{<tool, solution>}, treating each instance of a problem-solution pair or tool-solution pair as a unique connection. We utilized the \texttt{plotly} library~\cite{plotly} to generate the Sankey diagrams~\cite{Schmidt2008}, though it is important to note that the diagrams presented in this paper focus only on the most prominent connections for the sake of space and readability, while the complete set of diagrams encompasses all collected sources.

\subsection{Iterations}\label{sec:iterations}
As previously mentioned, we conducted three iterations of the described procedure. The first search was conducted in September 2020 while the second search was conducted in February 2023. These iterations have been documented in our previous published studies~\cite{2021-Ricca-NEXTA,Quatic2023}. The third iteration search was conducted in April 2024 and it is presented in this paper.
In each iteration, we started from the previously built taxonomies and refined them by adding new categories and values if needed.

\begin{table*}[t]

\caption{Taxonomy of test automation problems identified in the first~\cite{2021-Ricca-NEXTA}, second~\cite{Quatic2023} (in green), and third iteration (in light blue).}
\label{tab:problemsIII}

\centering
\scriptsize

\setlength{\tabcolsep}{10.3pt}
\renewcommand{\arraystretch}{1}

\begin{tabular}{lrrr|lrrr}

\toprule

\sc \textbf{Problem}  & \textbf{\cite{2021-Ricca-NEXTA}} & \textbf{\cite{Quatic2023}} & \textbf{\#} & \sc  \textbf{Problem}  & \textbf{\cite{2021-Ricca-NEXTA}} &  \textbf{\cite{Quatic2023}} & \textbf{\#}\\

\midrule 

\bf Test Planning  	&	22	&	32	&	52	&	\bf Test Execution  	  	&	71	&	103	&	118	\\
\quad Critical paths identification 	&	13	&	13	&	19	&	\quad Untested code 	&	29	&	34	&	40	\\
\quad Planning what to test 	&	7	&	9	&	11	&	\quad Flakiness 	&	18	&	22	&	22	\\
\quad Planning long release cycles 	&	2	&	2	&	2	&	\quad Slow execution time 	&	14	&	16	&	23	\\
\quad \cellcolor{green!25}Test process management 	&	 - 	&	8	&	20	&	\quad Useless test re-execution	 	&	4	&	7	&	9	\\
			&		&		&		&	\quad Scalability 	&	2	&	3	&	13	\\
\bf Test Design  	&	8	&	22	&	55	&	\quad Parallelization 	&	2	&	3	&	5	\\
\quad Programming skills required 	&	5	&	11	&	25	&	\quad Low user responsiveness 	&	1	&	1	&		1\\
\quad Domain knowledge required 	&	3	&	9	&	12	&	\quad Platform independence 	&	1	&	1		&	5	\\
\quad \cellcolor{green!25}Designing effective Testcases  	&	 - 	&	1	&	17	&	\quad &		&		&		\\
\quad \cellcolor{green!25}Adherence to coding standards   	&	 - 	&	1	&	1	&	 & &					&		\\
	&		&		&		&	\bf Test Closure	 		  	&	47	&	59	&	96	\\
\bf Test Authoring  	&	110	&	133	&	220	&	\quad Manual debugging overhead 	&	18	&	22	&	26	\\
\quad Manual code development 	&	52	&	58	&	108	&	\quad Costly result inspection 	&	10	&	11	&	23	\\
\quad Manual API test development 	&	7	&	11	& 13	&	\quad Visual analysis 	&	19	&	25	& 39		\\
\quad Manual data creation 	&	19	&	24	&	37	&	\quad \cellcolor{green!25}Data Quality 	&	 - 	&	1	&	1	\\
\quad Test object identification 	&	13	&	13	&	19	&	\quad \cellcolor{cyan!25}Code Quality 	&	 - 	&	 - 	&	7	\\
\quad Cross-platform testing 	&	10	&	15	&	26	&		&		&		&		\\
\quad Mimic geo-location testing 	&	1	&	1	&	2	&		&		&		&		\\
\quad Costly exploratory testing 	&	5	&	5	&	6	&	\bf Test Maintenance	 	    	&	82	&	102	&	184	\\
\quad Locators for highly dynamic elements		&	1	&	1	&	1	&	\quad Manual test code migration 	&	3	&	3	&	4	\\
\quad Test code modularity 	&	1	&	1	&	1	&	\quad Bug prediction 	&	11	&	13	&	17	\\
\quad Accessibility testing 	&	1	&	1	&	1	&	\quad Fragile test script 	&	10	&	12	&	27	\\
\quad \cellcolor{green!25}Adequacy-focus on faulty-areas 	&	 - 	&	3	&	5	&	\quad Regression faults 							&	2	&	7	&	7	\\
\quad \cellcolor{cyan!25}Optimize test strategies	&	 - 	&	 - 	&	2	&	\quad Costly visual GUI regression 		&	8	&	10	&	19	\\
	&		&		&		&	\quad Maintenance overhead 						&	48	&	57	&	100	\\
\bf Test Type  	&	 - 	&	 - 	&	3	&	\quad \cellcolor{cyan!25}Change Requirements	&	 - 	&	 - 	&	1	\\
\quad \cellcolor{cyan!25}Conducting thorough security testing	&	 - 	&	 - 	&	2	&	\quad \cellcolor{cyan!25}Communication developers-testers	&	 - 	&	 - 	&	5	\\
\quad \cellcolor{cyan!25}Testing-AI-generated-code	&	 - 	&	 - 	&	1	&	\quad \cellcolor{cyan!25}Troubleshooting and root cause analysis	&	 - 	&	 - 	&	4	\\
															
Unspecified 	&	60	&	77	&	 133	&	Generic 	&	23	&	33	&	55	\\
															
\midrule 															
															
Total  	&		&		&		&		&	423	&	545	&	917 \\

\bottomrule

\end{tabular}
\end{table*}

\section{Results}\label{sec:results}

\subsection{RQ\textsubscript{1} (Issues/problems)}

\autoref{tab:problemsIII} reports the existing TA problems and issues and the number of individual occurrences. 
The table reports the data collected in the first~\cite{2021-Ricca-NEXTA} and in the second~\cite{Quatic2023} (in green) iteration, as well as the updated version related to the third iteration (in light blue). 

Overall, in the three iterations, we identified 917 individual occurrences \changed{(545 after the second iteration, and 423 after the first one)} into the seven main categories: (1)~test planning, 5.6\% of occurrences; (2)~test design, 5.9\% of occurrences; (3)~test authoring, 24.1\% of occurrences; (4)~test type, 0.3\% of occurrences; (5)~test execution, 12.8\% of occurrences; (6)~test closure, 10.4\% of occurrences; and (7)~test maintenance, 20\% of occurrences. Other two categories such as Unspecified and Generic collected respectively 14.5\% and 5.9\% of occurrences. 

The most represented subcategory is \textit{Manual code development} (11.6\%), which was not surprising since it is well known~\cite{WCRE,Leotta2023} that the development of test cases and scripts is a complex task that requires non-trivial domain knowledge and appropriate testing and programming skills.

\begin{table*}[!ht]

\caption{Taxonomy of test automation solutions identified in the first~\cite{2021-Ricca-NEXTA}, second~\cite{Quatic2023} (in green), and third iteration (in light blue).}
\label{tab:solutionsIII}

\centering
\scriptsize

\setlength{\tabcolsep}{8.9pt}
\renewcommand{\arraystretch}{1}

\begin{tabular}{lrrr|lrrr}

\toprule

\sc \textbf{Solution}  & \textbf{\cite{2021-Ricca-NEXTA}} & \textbf{\cite{Quatic2023}} & \textbf{\#} & \sc  \textbf{Solution}  & \textbf{\cite{2021-Ricca-NEXTA}} &  \textbf{\cite{Quatic2023}} & \textbf{\#}\\

\midrule 

 \bf Test Generation  	&	 125 	&	192	&	302	&	 \bf Debugging  	&	 62 	&	82	&	149	\\
\quad Aut. test generation 	&	 29 	&	48	&	74	&	 \quad Intelligent test analytics 	&	 17 	&	22	&	27	\\
\qquad Aut. generation using machine translation 	&	 11 	&	13	&	25	&	 \quad Automated coverage report 	&	 14 	&	22	&	28	\\
\qquad Aut. generation from user behaviour 	&	 11 	&	22	&	29	&	 \quad Noticeable code changes identification 	&	 12 	&	13	&	23	\\
\qquad Aut. test generation from API calls 	&	 6 	&	6	&	13	&	 \quad Runtime monitoring 	&	 10 	&	11	&	12	\\
\qquad Aut. test generation from mockups 	&	 3 	&	3	&	3	&	 \quad Flaky test identification 	&	 7 	&	8	&	9	\\
 \qquad Aut. test generation using crawling 	&	 2 	&	11	&	16	&	 \quad Bad smell identification 	&	 1 	&	1		&	1\\
 \qquad \cellcolor{cyan!25}Aut. test generation using GenAI 	&	 - 	&	-	&	18	&	 \quad \cellcolor{cyan!25}Intelligent log analysis 	&	 - 	&	-		&	5\\
 \qquad \cellcolor{cyan!25}Codeless test generation 	&	 - 	&	-	&	11	&	 \quad \cellcolor{cyan!25}Intelligent test reporting 	&	 - 	&	-		&	6\\
 \quad \cellcolor{green!25}Declarative testing 	&	 - 	&	2	&	3	&	 \quad Decoupling test framework from host 	&	 1 	&	1	&	7	\\
 \quad \cellcolor{green!25}Predict faulty-areas 	&	 - 	&	4	&	4	&	 \quad \cellcolor{green!25}Root-cause-analysis 			&	1	&	10	&	10\\
 \quad Aut. data generation 	&	 22 	&	22	&	32	&	 \quad \cellcolor{green!25}Prediction of failures 	&	 - 	&	3	&	21	\\
 \quad Robust element localization 	&	 13 	&	13	&	14	&		&		&		&		\\
 \quad Dynamic user-behaviour properties recognition 	&	 8 	&	8	&	8	&	 \bf Maintenance  	&	 81 	&	141	&	201	\\
 \quad Automated exploratory testing 	&	 7 	&	10	&	12	&	 \quad Self-healing mechanisms 	&	 43 	&	43	&	63	\\
 \quad Object recognition engine	 	&	 6 	&	13	&	19	&	 \qquad Self-healing test scripts 	&	 24 	&	32	&	55	\\
 \quad Mock generation 	&	 3 	&	3	&	3	&	 \qquad Smart locators 	&	 19 	&	21	&	24	\\
 \quad Self-learning 	&	 2 	&	3	&	7	&	 \quad Intelligent fault prediction		&	 12 	&	13	&	13	\\
 \quad Automated API generation 	&	 1 	&	9	&	9	&	 \quad Intelligent selective test re-execution 	&	 12 	&	12	&	12	\\
 \quad Page object recognition 	&	 1 	&	2	&	2	&	  \quad Intelligent waiting sync 												&	 5 	&	5	&		5\\
 	&		&		&		&	 \quad Intelligent test prioritization								 		&	 4 	&	6	&	11	\\
	\bf Test Optimization		&	 - 	&	-	&	12	&	 \quad Aut. identification environment 		&	 \multirow{2}{*}{3} 	&	\multirow{2}{*}{6}	&	\multirow{2}{*}{6}	\\
	\quad \cellcolor{cyan!25}Improve test quality	&	 -	&	-	&	8	&		\quad			  configurations					&	  	&		&		\\
	\quad \cellcolor{cyan!25}Static (AI) Code Analysis	&	 - 	&	-	&	2	&	 \quad Pattern recognition 	&	 1 		&	1	&	3	\\
	\quad \cellcolor{cyan!25}Improve test scalability	&	 - 	&	-	&	1	&	 \quad Remove unnecessary test cases 		&	 1 	&	1	&	6	\\
	\quad \cellcolor{cyan!25}Anonymous test data	&	 - 	&	-	&	1	&	 \quad \cellcolor{green!25}Reduce UI testing  	&	 - 	&	1	&	3	\\
    	&	  	&		&										&	  	&	  		&		&		\\
  \bf Test Execution 	&	 - 	&	8	&	40	&	 \bf Test Process 		&	 - 	&	-	&	33	\\
  \quad \cellcolor{green!25}Cloud execution  	&	 - 	&	2	&	11	&	 \quad \cellcolor{cyan!25}AI data-driven test decisions  	&	 - 	&	-	&	13	\\
  \quad \cellcolor{green!25}Decoupling test framework from host  	&	 - 	&	2	&	8	&	\quad \cellcolor{cyan!25}AI (Chat-bots) for communication	&	-	&		-&	4	\\
  \quad \cellcolor{green!25}Smart test execution  	&	 - 	&	3	&	19	&	\quad \cellcolor{cyan!25}Shift Left Testing	&	-	&		-&	2	\\
  \quad \cellcolor{green!25}Anomaly detection 	&	 -	&	1	&	1	&	\quad \cellcolor{cyan!25}Hyper Automation Testing	&	-	&	-	&	1	\\
  \quad \cellcolor{cyan!25}Headless execution  &	- 	&	-	&	1	&		&		&		&		\\
		  &	 	&		&		&		\bf Test type	&	-	&		-&	4	\\
			&	 	&		&		&		\quad \cellcolor{cyan!25}IOT testing	&	-	&	-	&	2	\\
			&	 	&		&		&		\quad \cellcolor{cyan!25}Blockchain testing	&	-	&	-	&	2	\\
			&	 	&		&		&		&		&		&		\\														
 \bf Oracle  	&	 38 	&	46	&	70	&	 Unspecified 	&	 91 	&	99	&	119	\\
 \quad Visual testing 	&	 38 	&	46	&	70	&	 Generic 	&	 25 	&	30	&	55	\\
																		
\midrule 
					
\multicolumn{5}{l}{Total}  &	466	&	607	&	972	\\

\bottomrule

\end{tabular}
\end{table*}

The second most mentioned subcategory is related to \textit{Maintenance overhead} (10.9\%). 
Nowadays, software applications evolve continuously, in particular, in fields such as web and mobile. This continuous evolution of software applications requires continuous maintenance and evolution also of corresponding test suites, thus being aligned with the applications. 
Existing research~\cite{WCRE,Leotta2023} shows that test maintenance is a highly expensive and time-consuming task, and in some cases, it can even be the most costly test automation activity~\cite{Rafi2012BenefitsAL}.

Other representative subcategories, even if less numerous, are: \textit{Untested code} (4.3\%), \textit{Visual analysis} (4.2\%), and \textit{Manual data creation} (4\%).

\textit{Untested code} and \textit{Manual data creation} are key aspects to optimize the generation of effective test cases. On the one side, it is impossible to test everything in a software application, so strategies and techniques to identify the less tested portion of the application are fundamental for reducing the overall effort by focusing on where it is needed. For instance, it has been demonstrated that test suites with low coverage of the app code have a lower chance of detecting bugs~\cite{Brader-2016}.
On the other side, the use of high-quality test data is a critical part of testing~\cite{2021-Ricca-SOFSEM}, but producing such high-quality test data can be time-consuming.

Validating the visual correctness of a GUI (\textit{Visual analysis} subcategory) is a particularly challenging task. When done manually, testers must visually inspect all elements of the application to ensure they appear as intended, often across multiple devices and platforms. This process generally involves comparing screenshots of the current application against a previously established baseline, or ``golden master'', and reporting any significant visual discrepancies.

Another significant challenge in test automation is test flakiness (\textit{Flakiness subcategory}). A test script is considered flaky if its execution on the same application results in inconsistent outcomes due to environmental factors like screen size, browser version, or network conditions~\cite{leinen2024cost,10.1145/2635868.2635920}. This issue undermines the reliability of test automation, as flaky tests are more prone to missing defects (false negatives) or reporting incorrect errors (false positives).

\begin{table*}[t]

\caption{The most popular AI-based tools used in the context of software testing, as identified in the first~\cite{2021-Ricca-NEXTA}, second~\cite{Quatic2023} (in green), and third iteration (in light blue).}
\label{tab:tools}

\centering
\scriptsize

\setlength{\tabcolsep}{15.5pt}
\renewcommand{\arraystretch}{1}

\begin{tabular}{llrlrlr}

\toprule

& \multicolumn{2}{c}{\textbf{\cite{2021-Ricca-NEXTA}}} 
& \multicolumn{2}{c}{\textbf{\cite{Quatic2023}}}  
& \multicolumn{2}{c}{\textbf{Third Iteration}}  \\

\cmidrule(r){2-3}
\cmidrule(r){4-5}
\cmidrule(r){6-7}

& \multicolumn{1}{c}{\textbf{Tool}} 
& \multicolumn{1}{c}{\#}
& \multicolumn{1}{c}{\textbf{Tool}} 
& \multicolumn{1}{c}{\#}
& \multicolumn{1}{c}{\textbf{Tool}} 
& \multicolumn{1}{c}{\#} \\

\midrule 

1 & Testim & 14 & Applitools & 24 & Applitools & 50 \\
2 & Applitools & 12 & Testim & 20 & Testim & 38 \\
3 & Functionize & 11 & Functionize & 16 & Functionize & 35 \\
4 & Mabl & 7 & Mabl & 14 & AcceIQ & 32 \\
5 & LambdaTest & 6 & TestCraft & 9 & Mabl & 29 \\
6 & Laucnhable & 5 & Parasoft SOA test's Smart & 8 & TestCraft & 21 \\
  &   				 &   & \quad API Test Generator & 8 &   &  \\
7 & Parasoft SOA test's Smart & 5 & SauceLabs & 8 & Testsigma & 18 \\
  & \quad API Test Generator &  &  & 8 &  &  \\
8 & QMetry Digital 	 & 5 & TestComplete & 8 & Katalon & 15 \\
  & \quad Quality Platform & 5 &  &  &  &  \\
9 & Testsigma & 5 & UiPath Test Suite & 8 & \cellcolor{green!25}ChatGPT & 14 \\
10 & UiPath Test Suite & 5 & AcceIQ & 7 & TestComplete & 12 \\
11 & Test.AI & 4 & \cellcolor{green!25}ChatGPT & 7 & Parasoft SOA test's Smart  & 10 \\
   &   &  &  &  & \quad API Test Generator &  \\
12 & Tricentis Tosca & 4 & LambdaTest & 7 & UiPath Test Suite & 10 \\
13 & AcceIQ & 3 & Testsigma & 7 & LambdaTest & 9 \\
14 & Appium & 3 & Test.AI & 6 & SauceLabs & 9 \\
15 & Google OSS-Fuzz & 3 & Eggplant AI & 5 & \cellcolor{cyan!25}Aqua & 8 \\
16 & Kobiton & 3 & Katalon & 5 & Test.AI & 7 \\
17 & Percy & 3 & Laucnhable & 5 & \cellcolor{green!25}Testrigor & 7 \\
18 & SauceLabs & 3 & Parasoft Selenic & 5 & Appium & 6 \\
19 & TestCraft & 3 & QMetry Digital & 5 & Eggplant & 6 \\
 &  &  & \quad Quality Platform &  &  &  \\
20 & TestProject & 3 & Tricentis Tosca & 5 & Parasoft Selenic & 6 \\
21 & Appvance IQ & 2 & Appium & 4 & \cellcolor{cyan!25}pCloudy & 6 \\
22 & Browsershots & 2 & BrowserStack & 4 & Tricentis Tosca & 6 \\
\midrule 
& \multicolumn{2}{r}{\bf 50} & \multicolumn{2}{r}{\bf 67} & \multicolumn{2}{r}{\bf 100}\\

\bottomrule

\end{tabular}
\end{table*}

\subsection{RQ\textsubscript{2} (Solutions/approaches)}

\autoref{tab:solutionsIII} reports the existing AI-based solutions to the evidenced TA problems and issues (RQ\textsubscript{1}). The table reports the data collected in the first~\cite{2021-Ricca-NEXTA} and in the second~\cite{Quatic2023} (in green) iteration, as well as the updated version related to the third iteration (in light blue). 

The solutions have been grouped into eight main categories. In the table, TA solutions and their occurrences are listed for iterations. Overall, in the three iterations, we identified 972 individual occurrences \changed{(607 after the second iteration, and 466 after the first one)} into the eight main categories: (1)~test generation, 31\% of occurrences; (2)~test oracles, 7.2\% of occurrences; (3)~debugging, 15.3\% of occurrences; (4)~test maintenance, 20.6\% of occurrences; (5)~test process, 2\% of occurrences; (6)~test execution, 4.1\% of occurrences; (7)~test optimization, 1.2\% of occurrences; and (8)~test type, 0.4\% of occurrences. Other two categories such as Unspecified and Generic collected respectively 12.2\% and 5.6\% of occurrences.

The most represented subcategory is \textit{Automatic test generation} (189 occurrences out of 972, 19.4\%, by considering all automatic test generation subcategories). Automatic test case generation, however, remains an ambitious task, even when AI-based technologies are used. We could observe that 7.6\% of the occurrences related to automatic test generation did not give any indication about how automated test code generation is implemented. Existing works generate tests automatically by: (i)~starting from the analysis of the real users' behaviors (2.9\%); (ii)~by applying machine translation techniques (2.5\%), for example, natural language processing techniques that develop test scripts starting from test requirements descriptions written in natural language; and (iii)~recently, by adopting generative AI strategies (1.8\%).

The second most mentioned subcategory is related to \textit{Visual testing} (7.2\%). To implement visual testing strategies different computer vision solutions are applied to automatically identify functional and visual problems in the application GUI, in particular, by applying image-recognition and OCR techniques to identify graphical elements and detect changes among the different tested versions.

Another representative subcategory, even if less numerous, is related to test maintenance, i.e., \textit{Self-healing mechanisms} (6.4\%).
Self-healing refers to the capability of automatically applying corrective actions when a test script is broken, for instance, after an application maintenance or evolution task, without human intervention. The continuous evolution of modern applications breaks test scripts, thus requiring a large effort to repair such test scripts. 
Following the analysis of the documents, we further split this subcategory into two distinct categories: \textit{Self-healing test script} (5.6\%) and \textit{Smart Locators} (2.4\%). 
The adoption of smart locators can be relevant to prevent the need for repairing interventions. By using multiple constructs that can be updated dynamically as the application evolves smart locators tend to be resilient to test breakages. In detail, by using multiple attributes per web element to locate, smart locators improve the robustness of test scripts.

\subsection{RQ\textsubscript{3} (Tools and Platforms)}

\autoref{tab:tools} shows that we identified in total 100 tools that support the TA solutions strategies (RQ\textsubscript{2}) and that were discussed in the analyzed documents in the three iterations \changed{(67 after the second iteration, and 50 after the first one)}. Among these tools, the table lists the twenty-two most frequently cited and discussed ones. 
Considering the third iteration, the top-5 tools are: \textit{Applitools},\footnote{\url{https://applitools.com}} \textit{Testim},\footnote{\url{https://www.testim.io}} \textit{Functionize},\footnote{\url{https://www.functionize.com}} \textit{AccelQ},\footnote{\url{https://www.accelq.com}} and \textit{Mabl}.\footnote{\url{https://www.mabl.com}}

Applitools focuses on GUI testing by applying automated visual testing strategies based on different computer vision technologies and algorithms.
Testim applies intelligent capture-replay approaches and GPT-based technology for automatic test generation and adopts smart locators to prevent test breaking. Testim also supports AI data-driven testing decision strategies.
Functionize uses advanced NLP technologies for automatic test generation and adopts AI-based strategies for applying self-healing maintenance, in particular, for dynamically updating test scripts based on the application changes. 
Mabl uses a crawler for exploring a web application aiming at automatically generating test scripts by covering all reachable parts of the application under test. Mabl offers also a self-healing test script
solution and supports AI data-driven testing decision strategies.
Finally, AccelQ uses generative AI to automatically generate tests and offers self-healing test scripts, as well as an efficient cloud-based test execution.

\changed{Our survey extension highlighted a significant increase in the availability of new tools for developers. Their number has grown significantly in a short time, and new options have appeared on the market. We find ChatGPT, Aqua, TestRigor, and pCloudy among the new tools and platforms. ChatGPT, an advanced language model from OpenAI that generates human-like text, is primarily used for creating test scripts in software testing and was the most frequently mentioned tool in the documents we reviewed, with 14 total occurrences. Following that, we have Aqua, TestRigor, and pCloudy.
Aqua is a test management platform that leverages AI to optimize test planning, execution, and tracking. Offering cloud-based solutions for both manual and automated testing, it integrates with various tools to improve testing efficiency and effectiveness.
TestRigor is an automated testing tool designed for creating and running end-to-end tests using plain English commands. It simplifies test creation and maintenance.
pCloudy is a unified app testing platform that ensures app quality across various devices and browsers. It offers cloud-based manual and automated testing with access to thousands of real devices. Leveraging AI, pCloudy enhances integration, efficiency, and effectiveness in testing.}

\begin{figure*}[t]
\centering
\includegraphics[width=\linewidth]{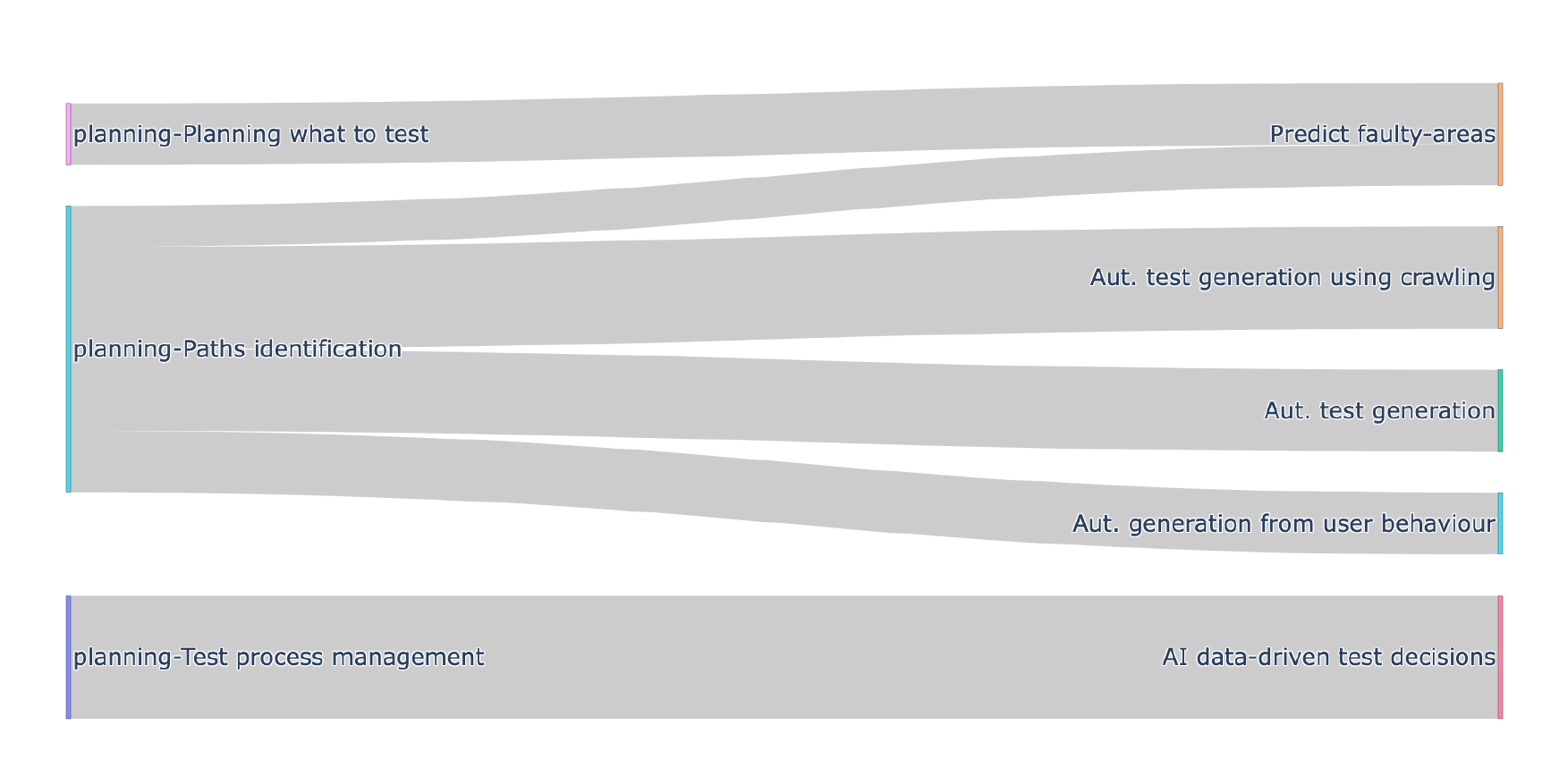} 
\caption{Problems vs Solutions: Test planning Sankey diagram.}
\label{fig:planning}
\end{figure*}

\begin{figure*}[t]
\centering
 \includegraphics[width=\linewidth]{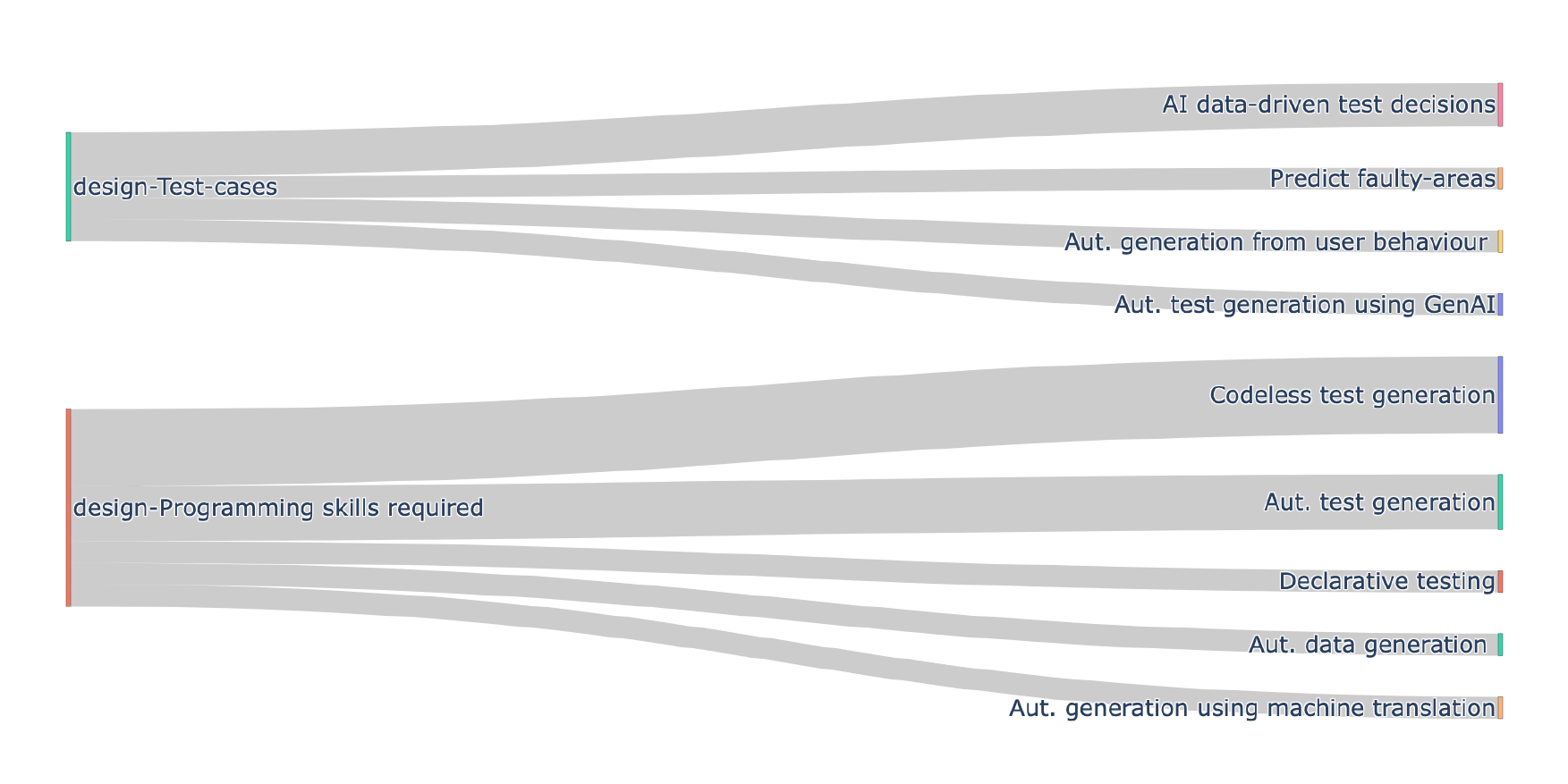}
\caption{Problems vs Solutions: Test design Sankey diagram.}
\label{fig:design}
\end{figure*}

\begin{figure*}[t]
\centering
\includegraphics[width=\linewidth]{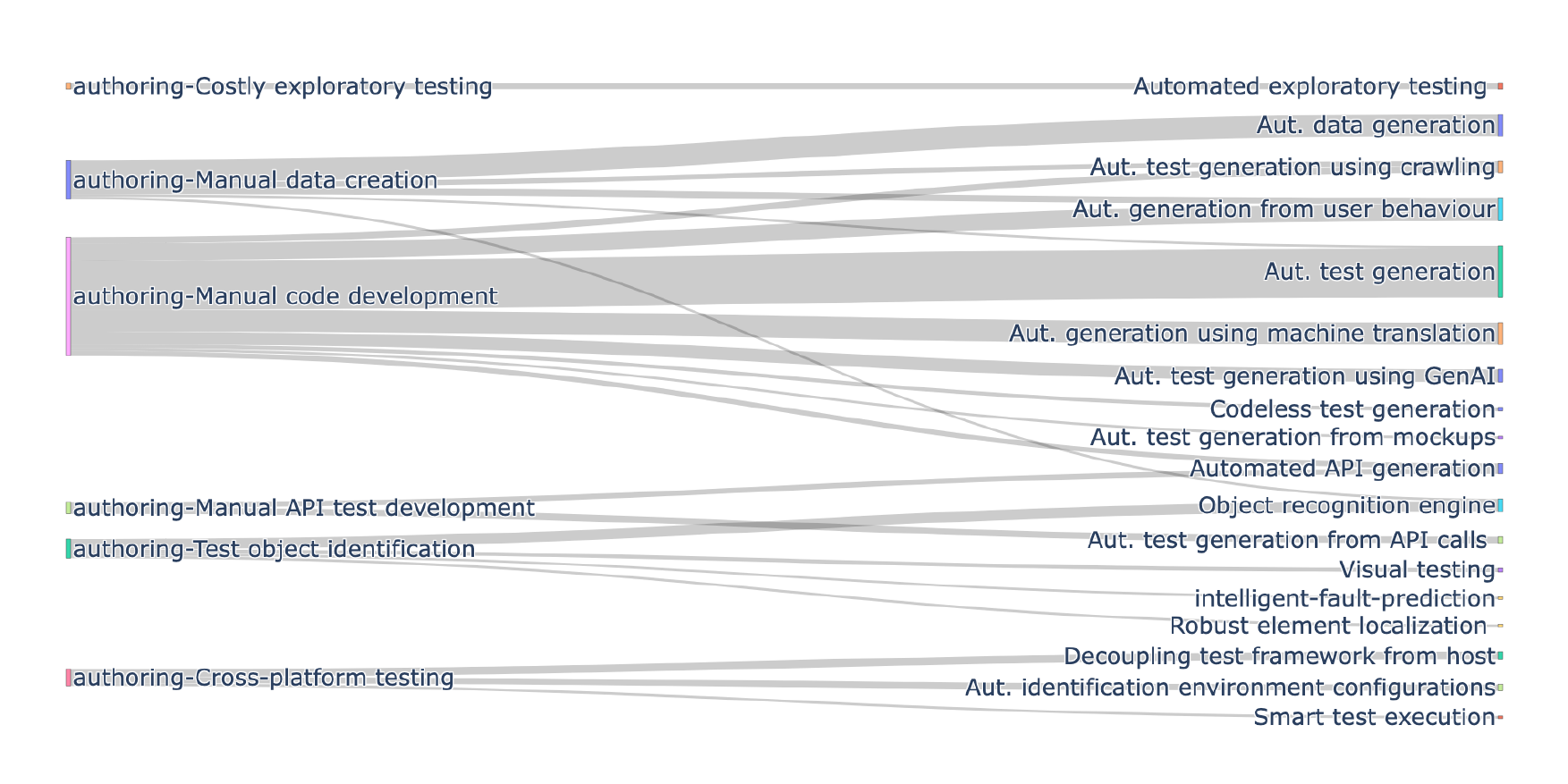}
\caption{Problems vs Solutions: Test authoring Sankey diagram.}
\label{fig:authoring}
\end{figure*}

\subsection{RQ\textsubscript{4} (Problems vs. Solutions)}

\autoref{fig:planning} utilizes a Sankey diagram to illustrate the relationships between problems and AI-based solutions in the test planning phase: problems are shown on the left, while solutions are shown on the right. 
In the planning phase, the most relevant problem (the larger node on the left) identified by the community is the identification of test paths, i.e., a specific sequence of actions that are executed to verify that a particular feature or functionality of the application under test is working as expected. This sequence can involve clicking on links, filling out forms, submitting data, and interacting with various elements on the web pages. In particular, testers are interested in identifying the critical paths in the app to test them thoroughly.
In terms of solutions adopted, automatic test generation based on crawling and user behaviors are the most frequently discussed. However, as shown in the diagram, other solutions for identifying critical paths are predicting faulty areas in the application under test and intelligent test analytics. 
It is worth noting that test process management has become an increasingly relevant issue over the years. In fact, there has been a noticeable rise in individual discussions on this topic within the community. As shown in \autoref{tab:problemsIII}, there were no occurrences of this issue identified in the first iteration, eight occurrences in the second iteration, and twenty occurrences in the most recent iteration. The most commonly used solution for this problem is using AI and data analytics to guide the software testing lifecycle. This approach leverages various technologies and methods to make informed decisions based on data rather than relying solely on manual processes or intuition.

Similarly, \autoref{fig:design} shows that the most relevant test problem when designing tests, concerns the programming skills required to develop test scripts (the larger node on the left), thus as expected, several different strategies and technologies are proposed (we can see in the diagram different output nodes for programming skills) for automatically generating test scripts and test data, e.g., the adoption of generative AI, NLP, codeless test scripts (e.g., by visual testing mechanisms). 
The second most important problem we can note is the design of test cases, which involves determining what to test, the steps to take to create a test case, and what the oracle is. We can see from the diagram that various solutions are proposed for this problem, ranging from using automated exploratory testing to automatic generation, for example by using generative AI.

\begin{figure*}[t]
\centering
\includegraphics[width=\linewidth]{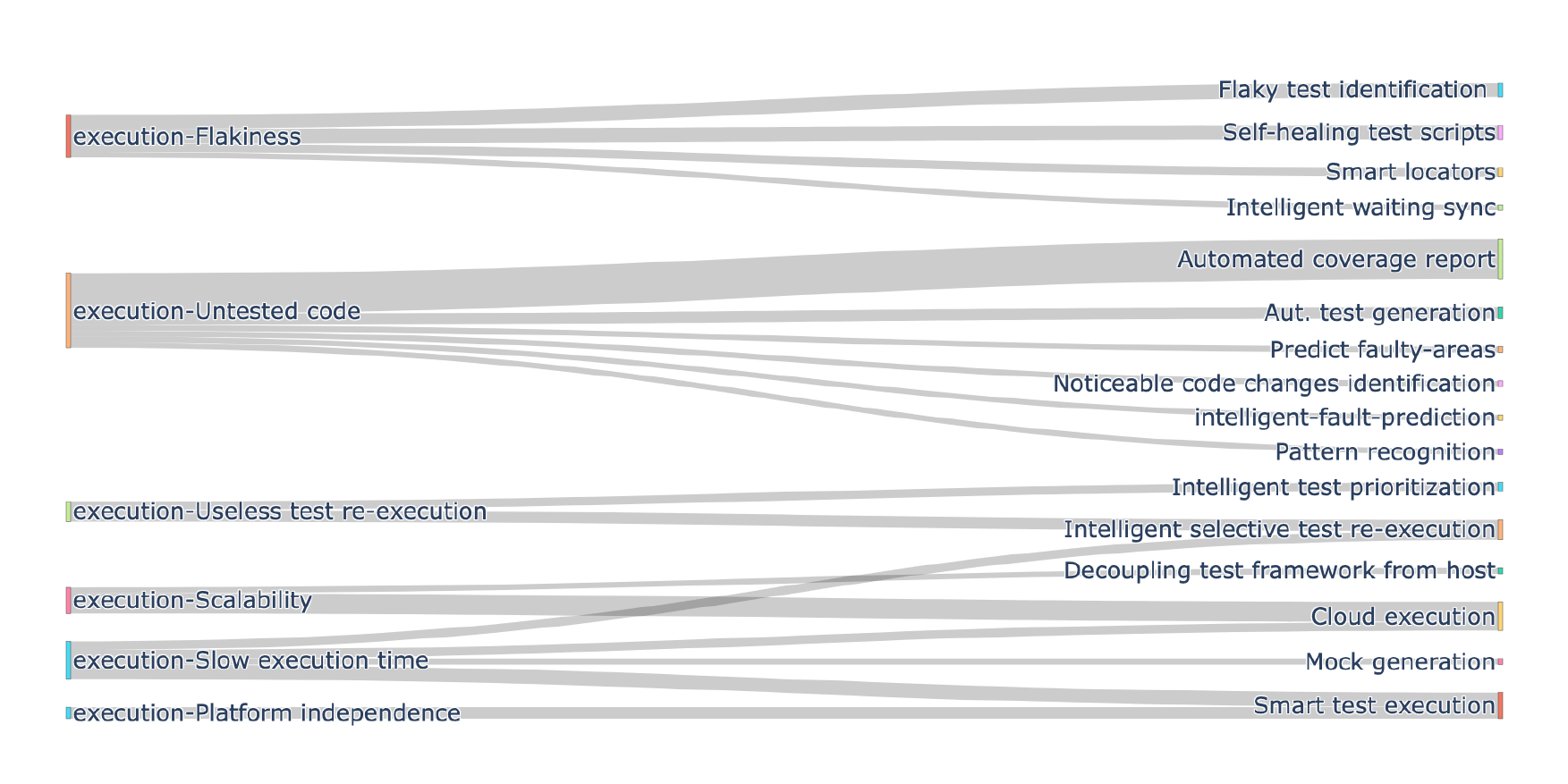}
\caption{Problems vs Solutions: Test execution Sankey diagram.}
\label{fig:execution}
\end{figure*}

\begin{figure*}[t]
\centering
\includegraphics[width=\linewidth]{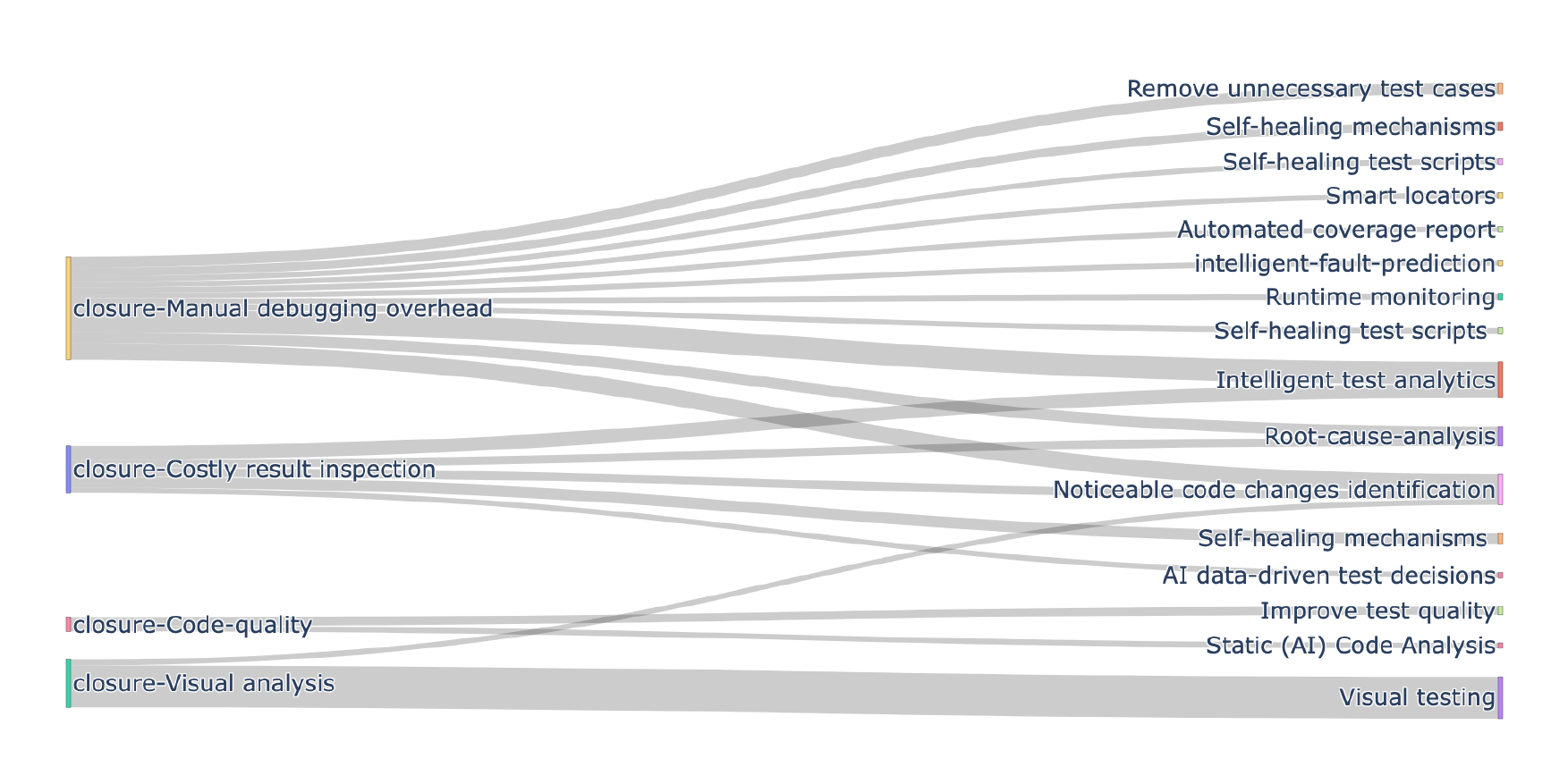} 
\caption{Problems vs Solutions: Test closure Sankey diagram.}
\label{fig:closure}
\end{figure*}

\autoref{fig:authoring} represents the Sankey diagram problems/solutions in the authoring phase and shows that the most relevant problems in this phase are related to the manual effort needed to develop/implement test cases in test scripts and create effective test data (see the two larger nodes on the left). 
Among the solutions to limit the manual effort in implementing test cases, we find that the predominant category, as we would expect, is automatic generation. The subcategories include all the automatic generation methods found, such as using screen mockups as a starting point, employing a web crawler, and even utilizing generative AI tools/platforms. Regarding the problem of generating effective test data, the main solution is to rely on AI-based tools capable of generating it.
It is interesting to note that test object identification---the ability of testing tools/platforms to correctly recognize and interact with various elements (objects) on a web page, e.g. dynamic or modal elements, and cross-platform problems---cross-platform testing involves verifying that a software application functions correctly across different environments, including various operating systems, browsers, and devices. Planning cross-platform testing presents unique challenges due to the diversity and complexity of environments on which the application must be tested---remain among the most relevant problems, in all the three considered iterations. 
In particular, the identification of test objects is faced with several strategies such as (i)~object recognition engines (i.e., identification of testing elements in the GUI); (ii)~robust element localization  (e.g., use of multiple sources to localize elements in the GUI); (iii)~visual testing (i.e., automated visual checks of the GUI by using of computer vision); and (iv)~intelligent fault prediction strategies.  

\begin{figure*}[t]
\centering
\includegraphics[width=\linewidth]{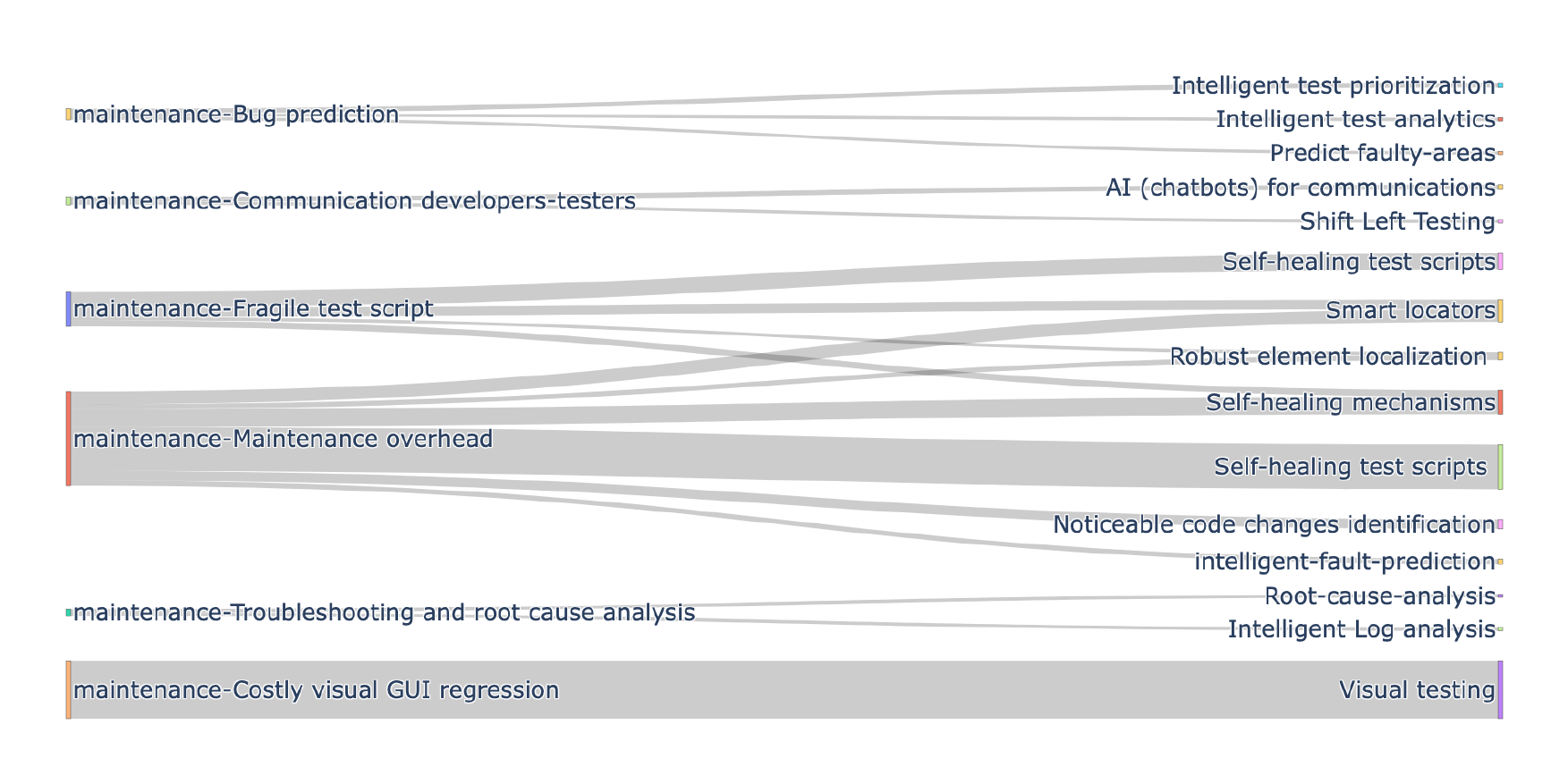} 
\caption{Problems vs Solutions: Test maintenance Sankey diagram.}
\label{fig:maintenance}
\end{figure*}

The Sankey diagram represented in \autoref{fig:execution} shows that several different problems seem to affect the test execution phase. 
Among the various problems, the most significant and those that attract the community's attention seem to be flakiness and untested code.
In the last iteration (see \autoref{tab:problemsIII}), flakiness---flakiness in tests occurs when test results are unreliable, showing varying outcomes across multiple runs even though the application state remains unchanged---appears to be recognized as less relevant compared to the first and second iteration. Conversely, the problem of identifying the untested code is becoming more and more relevant with the second and third iteration. 
A multitude of intelligent mechanisms (e.g., fault prediction, code change analysis, smart locators, API generation, exploratory testing, and test generation) are adopted to increase code coverage during testing.
In particular, the automatic generation of coverage reports is a key aspect for keeping developers and testers continuously informed about coverage information, enabling them to take compensatory actions if necessary.
Another notable issue in this phase is the lengthy execution time of test suites, which can be quite significant in some cases. The grey literature presents a variety of solutions for this problem, including 'smart' test execution approaches---smart test execution scans your application for code changes and runs tests specifically to validate those changes---, runtime monitoring, and the exclusion of non-essential test cases.

\begin{figure*}[t]
\centering
\includegraphics[width=\linewidth]{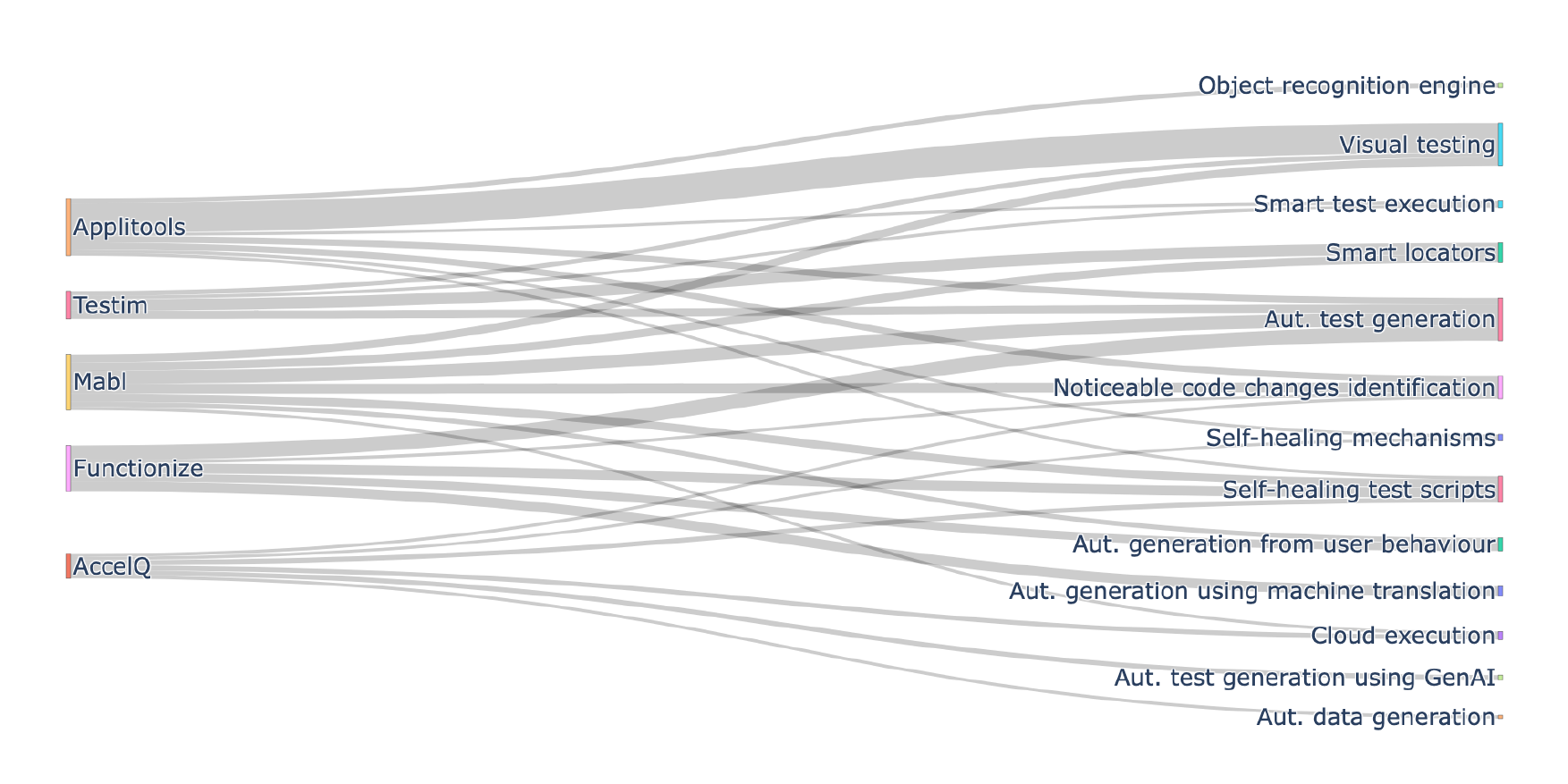} 
\caption{Tools vs Solutions: Sankey diagram.}
\label{fig:tools-vs-solutions}
\end{figure*}

\autoref{fig:closure} and \autoref{fig:maintenance} respectively depict the relevant problems and the adopted solutions for test closure and maintenance using Sankey diagrams.
Regarding test closure, the most relevant problem is still the manual overhead needed to debug the code. Two strategies can be identified among the provided solutions. On the one side, solutions such as root-cause analysis and automatic identification of code changes are proposed to identify and fix the issues by considering multiple information sources. On the other side, self-healing and intelligent analytic mechanisms are applied to avoid and predict issues, thus reducing the manual debugging overhead. Concerning test closure, in the three iterations, we also observed an increasing interest in techniques related to failure prediction and static code analysis, aiming at improving the code and data quality. 
The most significant problem in test maintenance is still the overhead involved. The strategies proposed in the solutions fall into three categories: (i)~the development of self-healing test scripts and mechanisms to avoid costly maintenance interventions; (ii)~the application of fault prediction and code change detection techniques to better target and minimize maintenance efforts; and (iii)~the implementation of smart locators and robust element localization methods to prevent test maintenance activities.
Other issues affecting test maintenance are automatically validating the visual correctness of the application and the fragility of tests. A solution to the problem of automatically validating the visual correctness of the application is automated visual verification of the web app GUI using computer vision techniques. Test fragility, defined as the tendency of tests to break easily due to minor changes in the application, can be addressed with various solutions, ranging from self-healing mechanisms to improving test script quality.

\subsection{RQ\textsubscript{5} (Tools vs. Solutions)}

\autoref{fig:tools-vs-solutions} presents a Sankey diagram featuring the five most frequently cited tools in the third iteration---Applitools, Testim, Mabl, Functionize, and AccelQ---as listed in \autoref{tab:tools}.
Each flow in the diagram represents a relationship identified in the analyzed documents between a tool (left node in the diagram) and the AI-based solutions (right node in the diagram) it supports, as listed in \autoref{tab:solutionsIII}.
We emphasize that, in our case, the size of the flows and nodes in the diagram indicates relationships that are more comprehensively addressed in the analyzed documents.
In the figure, it is evident that Functionize demonstrates notable strengths and diverse capabilities. Specifically, Functionize excels in using NLP and AI for automatic test generation and employs advanced techniques for self-healing test maintenance (see \autoref{fig:tools-vs-solutions}).
Similarly, Applitools provides flexibility and a comprehensive set of features. Notably, Applitools excels in visual testing and employs advanced strategies for automated test generation, including object recognition systems, user behavior observation, and generative AI mechanisms.
Applitools provides also test maintenance (e.g., self-healing mechanisms, code change identification capability) and debugging features (e.g., root-cause analysis, and intelligent test analytics).
In general, as we can also see from the figure, testing tools, and platforms are very flexible and support a wider range of solutions to different problems. However, there are exceptions, such as Appium and Testsigma (not present in the figure), which are more specialized and target a narrower set of capabilities.

\begin{figure*}[t]
\centering
\includegraphics[width=0.7\linewidth]{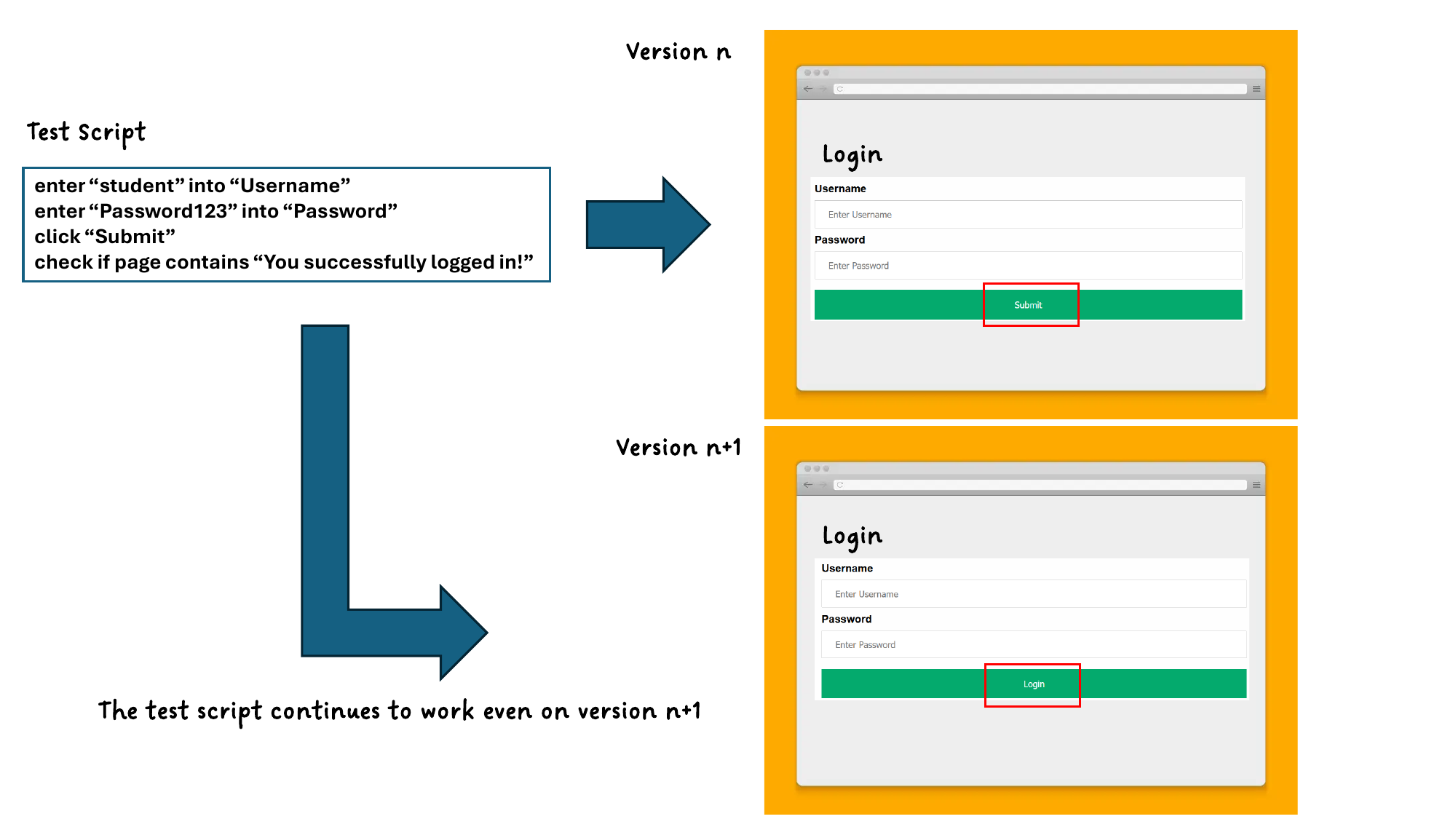}
\caption{Representation of the self-healing mechanism on the Login example.}
\label{fig:self-healing}
\end{figure*}

\section{Qualitative Examples: Automated Generation and Self-healing Test Scripts}

\changed{In this section, we exemplify three of the most popular AI-based TA features, aiming to provide a more concrete understanding of some of the AI solutions proposed in \autoref{tab:solutionsIII}: codeless test script generation, automated test script generation using GenAI and self-healing test scripts. We illustrate these features in the web domain using testRigor,\footnote{https://testrigor.com/} one of the most popular tools identified in the third iteration (see \autoref{tab:tools}). We selected testRigor because it offered a 14-day trial license for full use of the tool (although limited to a single test suite). It is important to emphasize that similar features are also implemented, perhaps in different ways, in other tools identified by our taxonomy.}

\changed{testRigor falls into the category of \emph{codeless testing tools}. It enables the creation of test scripts using plain English commands, eliminating the need for traditional programming. By leveraging advanced NLP algorithms, testRigor translates these plain English instructions into executable test scripts and executes them.}
\changed{testRigor is capable of interpreting and translating straightforward commands in test scripts, such as "open" to launch a webpage in a browser, "click" to interact with a button, "insert" to input text into a field, and "check" to validate the presence of specific text on a webpage. For example, a simple testRigor test to verify the successful log-in on a web application could look like this:}

\begin{verbatim}
enter "student" into "Username"
enter "Password123" into "Password"
click "Submit"
check if page contains "You successfully logged in!"
\end{verbatim}

\changed{In addition to being codeless, the testRigor tool leverages \emph{GenAI to automatically generate NLP-based test scripts} from informal natural language descriptions of a test case, simplifying the test development process. For instance, a description like "Test checkout process" enables the AI to generate a relevant test script description, reducing the need for manual input. Similarly, a more detailed description, such as "Develop a full test for adding a MacPro to a shopping cart", for a given e-commerce web app, allows testRigor to produce the following complete test script:}

\begin{verbatim}
Click on "search bar" at the top of the page  
Enter "MacBook" into "search bar"  
Press Enter   
Click on the first hyperlink containing "MacBook"  
Click on "Add to Cart"  
Click on "Go to Cart"  
Check that page contains "MacBook"  
Check that page contains "Quantity: 1"
\end{verbatim}

\changed{Although this approach minimizes manual effort, its accuracy and usefulness depend on the clarity and specificity of the descriptions provided.}

\changed{In addition to its generation features, testRigor also offers the ability to enable \emph{self-healing test scripts}. Self-healing in software engineering refers to systems that automatically detect and fix errors without human intervention. Through advanced algorithms and monitoring, these systems identify issues, take corrective actions, and restore functionality, minimizing downtime and improving reliability. In test automation, self-healing scripts can adapt to changes in the application under test (AUT) by automatically correcting themselves when a test fails.}

\changed{Imagine an automated test script designed to interact with a button on a web page or software interface by identifying it through a specific ID locator. This method functions seamlessly until a new version of the software alters the button’s ID. As a result, the test fails, unable to locate and interact with the button. Such failures lead to maintenance challenges that require developers or testers to update the test scripts with the new locator details. This process is often tedious and resource-intensive. Self-healing test automation addresses this issue by introducing adaptability directly into the testing framework. Instead of failing outright when encountering a modified element, such as a button with a changed ID, a self-healing system actively attempts to identify the element using alternative methods, such as analyzing attributes, context, and surrounding elements or using computer vision algorithms. This capability allows the test to continue running without human intervention, minimizing disruptions and significantly reducing the effort required for test maintenance.}

\changed{testRigor implements self-healing tests by leveraging a component called Vision AI to adapt to specification changes for rules and individual commands. Instead of failing, testRigor analyzes the screen to find alternative ways to achieve the intended actions. Vision AI uses computer vision to enable machines to interpret and understand visual elements of user interfaces---such as images, icons, buttons, and text---just like a human tester. This approach allows the test script to quickly adjust to changes in a web app, making it particularly effective for testing complex graphical user interfaces.}
\changed{To better understand the concept of self-healing test scripts and how this mechanism is implemented in testRigor, let us consider a web page containing a Login form (version n), as shown above in \autoref{fig:self-healing}. Since testRigor operates as a human tester would, we created a test script to perform the Sign-In action using the UI text as locators, as already shown above in the first test script example. Now, let us suppose that the application under test has been updated (version n+1), and the button web element has been changed to 'Submit' (as shown below in \autoref{fig:self-healing}).}

\begin{figure}[t]
\centering
\includegraphics[width=\linewidth]{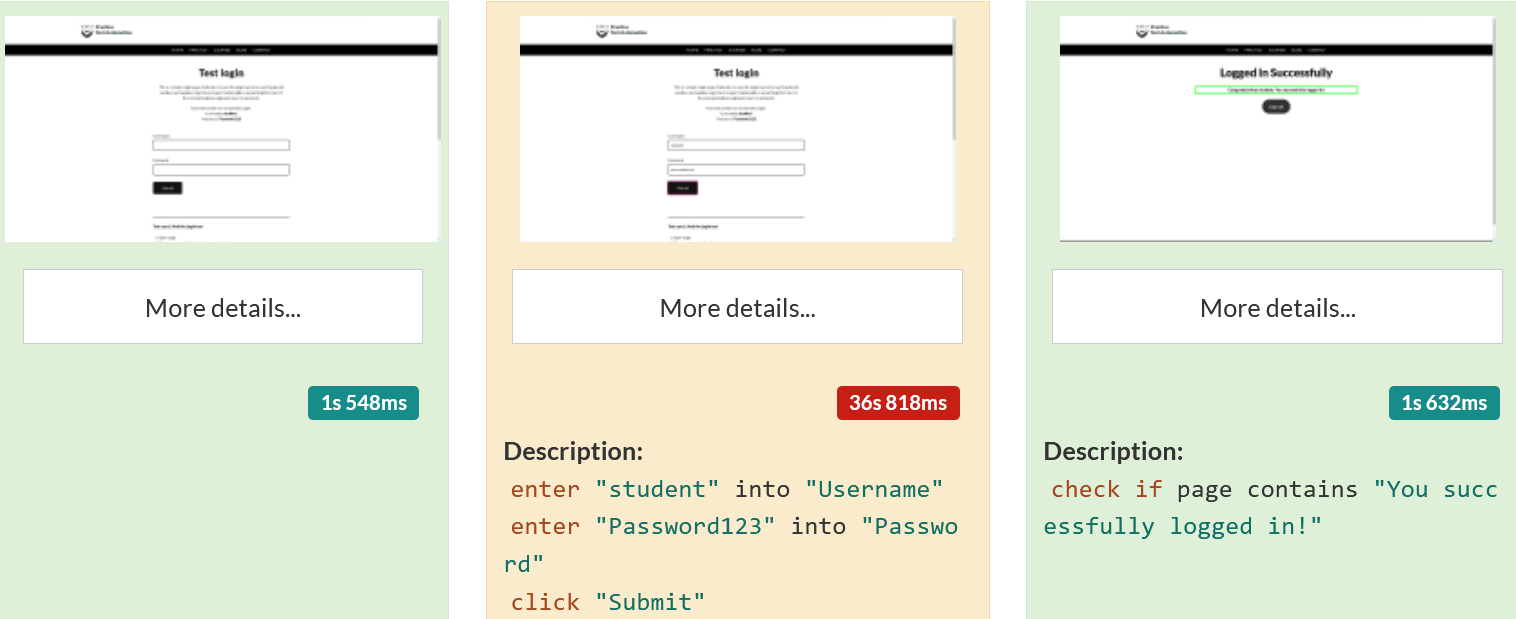} 
\caption{Test script execution with the self-healing mechanism enabled (represented by the box marked in orange).}
\label{fig:XYZ}
\end{figure}

\changed{A standard, non-self-healing test script would fail in this scenario because it would not locate the 'Login' text on the button. However, when executed with testRigor, which incorporates a self-healing feature, the test script successfully adapts and does not fail. As illustrated in \autoref{fig:XYZ}, the execution results show that the first step and the third step---corresponding to opening the webpage and executing the check command after clicking the Login button---are marked in green, indicating successful completion. Step two, however, is highlighted in orange, signaling that the self-healing mechanism was activated after entering the username and the password, to identify and adjust for the modified button in the updated version. Additionally, a link labeled ``More details...'' provides access to a detailed summary of the adjustments made.}
\changed{The test script execution demonstrates that when UI changes occur, testRigor’s AI attempts to identify alternative ways to interact with the affected element. By clicking the ``More details...'' button, a popup appears displaying the original test script alongside the AI-modified version with updated test steps. If reverting to the previous version is preferred, the ``Rollback to this'' option can be selected. Alternatively, if the AI-generated modifications are satisfactory, they can be retained as the final version.}

\section{Developers Interviews}\label{sec:interviews}

While grey literature provided valuable information for our study, the nature 
of these sources limits to suggestions on how to use AI for TA, without presenting concrete evidence of its practical application or usage details.  

To get a more complete picture, we have conducted semi-structured interviews with researchers and practitioners with various backgrounds and levels of expertise, focusing on the types of problems and solutions found in our taxonomies.

\subsection{Interviews Design}\label{sec:interviews-design}

\changed{The study focused on software engineers with experience in software testing and artificial intelligence. 
Convenience sampling was utilized to recruit participants, allowing for efficient enrollment within the study's constraints. However, this approach inherently carries a risk of bias due to the non-random selection process. To address this limitation, we encouraged word-of-mouth promotion of the interview opportunity, leading to the recommendation of diverse and appropriate candidates.
Only individuals with a minimum of five years of professional experience in software engineering were included. This criterion ensured that participants had sufficient exposure to both software testing and AI practices. Efforts were made to select participants representing a range of ages, genders, and professional roles. This approach aimed to incorporate diverse perspectives and reduce the risk of homogeneity in viewpoints.
Ultimately, five developers/researchers agreed to participate in our study and conducted semi-structured interviews to understand whether they perceive AI as a valuable asset that aligns with their needs and objectives in their testing activities.} 

\autoref{tab:interviews} provides details about the interviewees, including their position, expertise, and years of experience. Among the interviewed candidates the lowest value is 6 years and the highest is
20 years (median=14). Our pool includes professionals \changed{from a company in Sweden} involved in strategic decision-making, as well as technical personnel and a researcher \changed{at an applied science university in Austria with substantial experience in the industry.}

We conducted semi-structured interviews~\cite{799955}, integrating open-ended questions to gather unexpected insights and specific questions to maintain focus and assist interviewees.
After obtaining background information on the interviewees' general and AI-specific experiences, we began our questioning. The first question was deliberately broad: ``\textit{What types of AI have you been using in your work?}''. This approach aimed to introduce the topic openly, encouraging interviewees to discuss their experiences without steering them toward any particular problems/solutions. 

We then proceeded to more specific questions covering a wide range of topics in test automation, including test creation, maintenance, execution, and the tools used. We asked interviewees if they had encountered any issues or problems related to these topics. If they responded affirmatively, we provided more detailed questions to better understand the discussed topic.

All interviews were conducted remotely via Teams video calls, each lasting approximately 60 minutes. The interviews were transcribed using Teams' automated speech recognition tool, which converts audio/video files into text. After generating the automated transcriptions, we reviewed and manually corrected transcription errors. Finally, we proceeded with open coding of the transcribed interviews in which different parts of the transcribed text were tagged. \changed{The sanitized text is available in our replication package~\cite{replication-package}.}

\begin{table}[t]

\caption{Interview participants details.}
\label{tab:interviews}

\centering
\scriptsize

\setlength{\tabcolsep}{3.9pt}
\renewcommand{\arraystretch}{1}

\begin{tabular}{lllr}

\toprule

\textbf{Participant} 
& \textbf{\# Position}  
& \textbf{\# Expertise} 
& \textbf{\# Experience}
\\

\midrule 

Id1 & Q/A Manager & Q/A Strategy & 17+ years \\ 
Id2 & Senior Developer & Quality Monitoring & 6+ years \\ 
Id3 & Senior Developer & Data Science/LLMs &  10+ years \\ 
Id4 & Full Stack Developer & TA/DevOps & 14+ years \\ 
Id5 & Researcher & QA/Testing & 20+ years \\ 

\bottomrule

\end{tabular}
\end{table} 

\subsection{Interviews Results}\label{sec:interviews-results}

\subsubsection{Problems}

The usage of AI touched on several aspects of the software development process. For example, it is used for requirement elicitation during interviews with clients and customers to complement information and write summary reports. 

The participants highlighted the use of AI to enhance test execution, specifically through test prioritization and AI-assisted test refactoring. 
On a broader picture, AI was leveraged to analyze existing codebases for predictive maintenance and risk assessment. This approach helped reduce technical debt and maintenance challenges caused by the low skill levels of consultants that exhibit high turnover.

In end-to-end testing, the interviewees mentioned that a primary challenge is testability. The lengthy and complex nature of E2E test cases makes testing at the API level difficult, especially without a clear specification. E2E tests are often fragile due to the lack of a clear API, static checking, synchronization issues, and missing types. Creating a model for these tests is considered a valuable investment, as it can result in more maintainable and robust tests, especially if the web app model creation and maintenance can also be automated.

\subsubsection{Solutions}
The participants emphasized the use of AI to identify dependencies between services and detect obsolete test cases due to the lack of domain knowledge in complex codebases. Another proposed solution is to implement anomaly detection using test execution logs to monitor the system and production data to help anticipate failures before they occur.
An interesting case was the usage of LLMs for mutation testing: when new test cases were introduced in the test suite, the web app was modified with artificially generated faults to assess the quality of the evolved test suite.

Regarding self-healing in practice, the participants were quite critical, noting that not all issues can be ``self-healed''. They suggested limiting self-healing to bug classes that can be precisely described, otherwise bugs can go unnoticed. Some changes, such as updated IDs, can be managed with hash maps to fix mappings automatically or by rewriting page objects if there are changes. 

\changed{Ultimately, the participants see AI as a collaborative partner in testing, code review, and pair programming. With accountability resting on the developers, they write the tests (serving as specifications), while the AI generates the code. The usage of ``\textit{AI to create test cases}'' was reported to be ``\textit{in the very early stages at the moment}''. It was mentioned that AI was used to create API test mockups that developers complement manual testing by adding logic.}

The final message was to use AI to generate boilerplate code while keeping decision-making a human task.

\subsubsection{Tools}

Our participants highlighted their reliance on in-house tools and existing frameworks like Microsoft Azure DevOps, over individual tools or custom code bases that are hard to adapt to each specific customer's need. They also noted the integration of advanced generative AI models, such as OpenAI's ChatGPT and Microsoft's Copilot, into their daily activities for code generation and review.

The interviewees described employing a diverse collection of AI algorithms in their work. These range from traditional machine learning algorithms, such as Support Vector Machines (SVM) and decision trees. An interview mentioned ``conventional AI'' strategies such as search-based optimization algorithms, model-based approaches such as user interaction flows, or Markov Chain Monte Carlo for prioritizing tests and tracking the frequency of test changes.
The participants also mentioned the usage of computer vision techniques like object recognition to find web elements that are difficult to find in the DOM, internationalization (i18n) testing, or automating the visual oracle in tests.

\section{Threats to Validity}\label{sec:ttv}

This section discusses the limitations of this research and the validity of the results presented. The primary issues related to the validity of this grey literature review involve inaccuracies in data extraction, an incomplete set of studies due to the limitations of search terms and search engines, and potential researcher bias concerning the criteria for study inclusion and exclusion. In this section, we discuss four types of validity threats plus reproducibility according to a standard validity checklist~\cite{wohlin00}. 

Threats to \textit{internal validity} relate to potential biases and errors in the selection of documents (exclusion criteria) and the classification of the considered items. The classification task is particularly challenging in the context of grey literature because web documents are often informative rather than technical, and the terminology used can be vague and ambiguous. To mitigate as much as possible classification errors, we adhered to a systematic and structured procedure (Section 3.1) with multiple iterations, starting with a small pilot study for each iteration and ensuring a consistent approach.
Another potential threat to the internal validity of our grey literature review is the search string used in the review process. Modifications to the search string could lead to different results, highlighting a limitation of our approach. The search string we used was carefully designed, but any changes to it could yield different sets of documents, thus affecting the scope and findings of the review. This underscores the importance of a well-defined search strategy and acknowledges that our conclusions are based on the specific search parameters we employed.

The \textit{external validity} of our study is primarily limited by our selection of sources. We considered only documents available on Google within a specific time frame. Consequently, our findings may not be generalizable to documents from other search engines, repositories, or different time periods. Future research should expand the scope to include additional sources and broader time frames to validate and extend our findings. However, given the number of documents analyzed and considering that the analysis was conducted at different times in three separate iterations, we are quite confident in our results. 

\changed{Another potential threat to validity pertains to the interviews and the open coding process. While these were conducted solely by the third author, his prior experience with similar studies and analyses~\cite{humbatova2020taxonomy} lends credibility to the approach. The structured interviews were limited to five participants, a constraint largely due to the difficulty of engaging developers actively engaged in their professional responsibilities. Despite the small sample size, it is important to highlight that all participants were senior developers. This ensures a solid foundation for the insights gathered, given the depth of experience and expertise they brought to the discussions. Although larger or more diverse samples could potentially lead to different conclusions, our analysis of the open-ended responses revealed recurring themes and no emergence of new ones, suggesting that saturation was reached. Consequently, we believe that, despite the limitations, the sample size is adequate for this study. The use of convenience sampling could lead to a lack of representativeness in the sample, potentially skewing findings. To mitigate this, efforts were made to diversify the sample as much as possible within the method's constraints. However, it is acknowledged that the findings may not be fully generalizable to the broader population of software engineers.}

\textit{Construct validity} concerns the extent to which our methodology accurately captures the constructs of interest, such as the identification of AI-based solutions for test automation and their associated issues. 
The inherent vagueness and ambiguity in grey literature terminology pose a risk to construct validity. To address this, we defined clear criteria for inclusion and exclusion and employed a rigorous procedure with reliability checks through an initial pilot study conducted at each iteration.

\textit{Conclusion validity} pertains to the degree to which our conclusions are credible and dependable. The iterative nature of our experiment, along with a pilot study conducted at each iteration, helps to ensure that our conclusions are based on systematically derived evidence. Nonetheless, the subjective nature of document interpretation in grey literature reviews remains a potential threat.

To enhance \textit{reproducibility}, we have made all our results, including data, plots, and references, available in our replication package~\cite{replication-package}. This transparency allows other researchers to verify our findings and procedure. However, the inherent variability in grey literature sources means that exact replication may be challenging. We encourage future researchers to apply our procedure in different contexts to assess its robustness and adaptability.

\section{Discussion}\label{sec:discussion}

\subsection{Observations from the Survey} 

\subsubsection{Generalizability}
In the first iteration, we analyzed 156 documents (see \autoref{tab:data1}), in the second one we additionally considered 95 documents (+60.8\% to the initial set of 156 documents of~\cite{2021-Ricca-NEXTA}), and in this third iteration, we additionally considered 91 documents (+36.2\% concerning the total number of documents, that is, 251 found in the first two iterations).
Concerning the two initial taxonomies, in detail, we have only partially extended the taxonomy related to TA problems, keeping it largely unchanged from the original version~\cite{2021-Ricca-NEXTA}. 
In fact, during the second iteration, we only introduced 5 new subcategories of TA problems, resulting in a total of 44 individual occurrences out of 545. In the third iteration, we added 7 additional subcategories, which accounted for 25 individual occurrences out of 917.
We also made only partial progress in extending the taxonomy related to AI-based solutions. Specifically, during the second iteration, we introduced one new category (i.e., Test Execution) and 9 subcategories, which accounted for 80 individual occurrences out of 607. In the third iteration, we added 3 additional categories (i.e., Test Optimization, Test Process, and Test Type) along with 15 subcategories, resulting in 77 individual occurrences out of 972.
In conclusion, regarding the generalizability of the taxonomies, we can state that in our analysis, the list of problems and solutions has remained largely consistent with the previous versions, with only a few minor additions. On the other hand, the list of tools has shown significant and consistent growth across the iterations, doubling their number from the first to the third iteration.

\subsubsection{Prevalence of Web E2E Testing}

\changed{Our search query was designed to encompass software testing broadly, without focusing on any specific type of system. Nevertheless, our survey of the grey literature revealed that most articles centered on E2E testing for mobile and web applications, followed by unit and API testing.}

\subsubsection{AI for Innovative Test Authoring}
 
We identified TA problems for which AI-enhanced existing solutions seem to be promising and largely studied. In particular, we observed that test authoring (i.e., test creation) is the most investigated TA problem. We identified 34 tools (e.g, Applitools, Functionize, Testim, Testrigor) that use AI-based technology for automatically generating test cases using different technologies, such as NLP techniques or GUI-based capture and reply techniques to generate test scripts, aiming at limiting the manual human intervention. \changed{Research approaches have been proposed to this aim~\cite{2020-Biagiola-ICST,9671312}.}
Practitioners can use our taxonomy to easily identify the problems better supported by existing tools whereas researchers can identify unexplored problems for which they can propose innovative solutions. 

\subsubsection{AI-assisted Visual Oracle}

\changed{We identified TA problems that could be faced by AI-enhanced existing solutions implemented in AI-based tools. For instance, a well-known challenge in GUI testing is the creation of effective oracles~\cite{Oracles,gunel-issta-oracle,gunel-issta-tse}. The use of oracles based on visual testing, using AI and computer vision approaches, is suggested in the grey literature as one of the possible ways to face this challenge. This aligns with recent research visual approach to web testing~\cite{Stocco-2018-FSE,10795003}.}
Furthermore, 13 tools that support oracle visual testing have been identified, e.g., Applitools, AI Testbot, Mabl, Sealights, Testim, and Test.ai. Practitioners can use our findings to quickly select the most appropriate solutions for their TA problems and the most adequate tools that support the solutions to their TA problems.

\subsubsection{Phase-specific and Multi-Phase Solutions for Smarter Test Automation}

In general, we identified tools that support solutions across various phases of test automation and address multiple test automation problems. However, there are also, though less common, tools that are \textit{specific} to a particular phase or problem.
For instance, tools such as Applitools, Functionize, and Mabl can support different TA phases, e.g., test creation, maintenance, and execution. Conversely, tools such as Appium, Testsigma, and TestComplete seem to be more specific, thus mainly supporting a given testing phase, e.g., Appium and Testsigma focus on test creation, while TestComplete focuses on visual testing. Our findings are of interest to practitioners in selecting the most appropriate tools to use in their business, by taking into account the problems they have to face and also other aspects such as the specificity and flexibility of tools.

We also identified TA problems that require solutions involving \textit{several phases} of TA. For instance, in the test planning phase, we defined problems such as planning what to test, identifying critical paths in the application under test, and managing the entire test process, among others. 
\changed{We observe that the latter problem raised attention only recently (in the third iteration of our study), and it is faced by mainly applying AI data-driven decision strategies~\cite{8705808}. The second problem is mainly faced with automatic test creation solutions~\cite{2019-Biagiola-FSE-Diversity}, ranging from crawling the application under test~\cite{2020-Biagiola-ICST} to the creation of test scripts using user behaviors~\cite{9671312}.} The first problem instead is faced by solutions involving test script creation (e.g., automatic test creations by focusing on application areas that are predicted as more buggy), test selection (e.g., based on fault prediction), test execution (e.g., adoption of intelligent test re-execution strategies), and, finally, test debugging approaches (e.g., adoption of test analytics). 
Researchers can use our findings to better highlight as some TA problems are addressed from different perspectives, i.e., for some problems, specific ad-hoc solutions can be adequate while, for other problems, more complex solutions need to be studied. 

\subsubsection{Open Challenges for AIaTA}

\changed{We identified solutions presented in the grey literature that are not supported by existing available tools. For instance, nevertheless, test generation with mockups and dynamic properties identified from observed user behaviors are listed among possible solutions supported by AI, for automatically generating tests, we did not identify any existing tools that provided such capabilities. Non-AI research tooling has been proposed in literature~\cite{d2018enabling}, but AI-driven solutions remain limited, with only a few proposals, such as those leveraging deep learning~\cite{samir2024model}, introduced to date.}
However, it is important to keep in mind that the absence of a tool does not necessarily indicate a lack of existing solutions for a particular problem. It could simply mean that the specific tooling solutions were not mentioned in the literature due to the incompleteness of our analysis. For instance, decoupling the test framework from the host environment is referenced, in the grey literature, as one possible solution for facilitating cross-platform testing. 
    
It seems that this solution is not adequately supported by the existing AI-enhanced tools, that mainly provide approaches that allow the identification of different environmental configurations, to face cross-platform testing. \changed{Conversely, this issue has been addressed in the white literature for web~\cite{yu2021layout} and mobile domains~\cite{yu2021layout} with approaches based on robotic arms~\cite{xie2023nicropurelyvisionbasednonintrusive}, deep reinforcement learning~\cite{10299943}, and computer vision to enable the Chouette Crawler to help accelerate the mobile and web app testing cycle at Duolingo~\cite{10298753}.}
    
Another solution that seems to be not adequately supported by tools concerns the execution of test cases with mock responses: no tools support the construction of mock objects that can be used in TA. Concerning the test selection and optimization, solutions aiming at prioritizing test cases, removing unnecessary test cases and GUI-based testing seem to be not adequately supported by existing AI-enhanced tools. Our findings are of interest to both professionals and researchers to develop innovative tools and technologies capable of supporting the identified solutions. 

\subsection{Observations from the Interviews} 

\changed{We questioned whether AI is merely a more sophisticated form of traditional automation or a transformative technology that is here to stay. The consensus was largely positive, particularly regarding large language models. However, these tools require strict guidelines for junior developers, who might overly depend on their outputs without proper verification. Indeed, LLMs are prone to hallucinations, where the output appears plausible but is largely incorrect.
For test data generation, the creativity aspect of large language models can create interesting and useful test data in situations where optimization methods like fuzzing are not applicable. However, a clear limitation is AI's lack of domain knowledge. While AI can access vast amounts of data, it lacks specific domain or company-specific knowledge, making the pre-trained models insufficient.}

\changed{As such, our participants emphasized developing self-check mechanisms and best practices for using AI in TA. LLMs often attempt to create solutions rather than acknowledging uncertainty.  
LLMs have a user-pleasing tendency, finding workarounds instead of reporting errors (obviating the oracle problem). This can lead to excessively positive results, while testers typically adopt a more cautious, critical, and pessimistic perspective. Our interviews revealed that quality is not just about coverage; relevance and better assertions are crucial to avoid potential risks in production.
In practice, the testing team creates test suites exhibiting ``\textit{a reasonable coverage for their needs and for their criticality}''.}

\changed{Other research areas that deserve attention include human-computer interaction, focusing on ensuring that the results of AI are clear and explainable. Additionally, developing quality metrics to assess the output of AI models is essential, such as measuring the frequency of hallucinations accurately or evaluating the quality of the output they produce.}

\section{Related Work}\label{sec:relatedworks}

This section of related works is divided into three subsections: 1) works where AI and ML are applied to TA, 2) analyses of grey literature in the context of test automation and multivocal literature reviews, and 3) secondary studies focused on AIaTA.

\subsection{Artificial Intelligence in Test Automation}

In the software testing community, AI/ML solutions are increasingly adopted to automate various testing activities (e.g., test data generation) and address issues such as test suite maintenance and test case prioritization, ushering in a new era of smarter and more efficient QA processes.
Test generation is a significant area where AI/ML has been explored. Zhang et al.~\cite{Zhang-2017-ASE} and Walia et al.~\cite{Walia2022} propose, using computer vision, approaches to automate GUI test generation, aiming to reduce human effort. Qian et al.~\cite{Qian2023} adopt an OCR-based technique to localize GUI elements for test generation.
Test maintenance, which traditionally requires significant human effort (e.g., for page object generation~\cite{2017-Stocco-SQJ}), can also benefit from AI/ML. Computer vision approaches~\cite{2020-Bajammal-TSE} have been widely used for web test migration~\cite{2018-Leotta-STVR,2015-Leotta-SAC,2014-Stocco-SCAM} and test repair. Code-less functional test automation is investigated by Vos et al.~\cite{Vos2021} and Phuc Nguyen et al.~\cite{phucnguyen:hal-02909787} for test maintenance. The latter paper combines Selenium and ML techniques to reduce the time testers spend modifying test code.
Other testing issues addressed by AI/ML include ML-based detection of flaky tests (Camara et al.~\cite{Camara2021}), cross-browser incompatibility detection (Mahajan et al.~\cite{7102586}), test case prioritization (Feng et al.~\cite{Feng:2016:MTR:2970276.2970367}), and identifying areas of the application under test for increased test coverage such as the work by Yadav et al.~\cite{9698727}.
Unlike these works, our study does not focus on a specific AI/ML technique for test automation. Instead, we conduct a multi-year grey literature review to capture the state-of-the-art concerning test automation problems, proposed solutions, and existing tools.

\subsection{Grey Literature and Multivocal Literature Reviews on Test Automation}

One of the early works in grey literature analysis on test automation is that of Päivi Raulamo-Jurvanen and colleagues~\cite{10.1145/3084226.3084252}.
In this paper, the researchers investigate how practitioners address the challenge of selecting the appropriate test automation tool. Their methodology involves consolidating insights from practitioners through a review of grey literature sourced from 53 distinct companies. The findings reveal a shared understanding of important selection criteria, albeit with inconsistent application. To address this, the authors distill insights from various sources into a cohesive 12-step process and identify 14 distinct criteria for effective tool selection. The study indicates that practitioners generally exhibit a keen interest in and are influenced by related grey literature, as evidenced by the substantial number of backlinks to the sources. Despite the abundance of available software testing tools, practitioners tend to gravitate towards well-known and widely adopted options (e.g., Selenium, QTP/UFT, and TestComple). This work, although its objective differs from ours, shares similarities.

Another work falling into this category is that of Yuqing Wang et. al.~\cite{Wang_2022}. In their paper, the authors present a multivocal literature review aimed at surveying and synthesizing guidelines from existing literature on enhancing test automation maturity. A multivocal literature review is a type of systematic literature review that includes both academic literature and grey literature. They conducted a review of 81 primary studies. From these studies, they extracted 26 test automation best practices and collected various pieces of advice on how to implement them effectively. The main observations include: (1) Only 6 best practices have been empirically evaluated for their positive impact on maturity improvement; (2) Some technical best practices identified in the review were not previously included in test maturity models; (3) Certain best practices correlate with success factors and maturity impediments identified by other researchers; (4) Many pieces of advice on implementing best practices are derived from experience studies and require further empirical evaluation using formal methods; (5) Some advice on implementing certain best practices conflicts in the literature. The objective of this work is completely different from ours, and also the research method is different (multivocal literature review vs. grey literature analysis).

The last work we have selected in this category is that of Garousi and Mäntylä~\cite{GAROUSI201692}. The study investigates decision-making in software test automation within the context of software development and tries to answer the question: when and what to automate in software testing? While many organizations view test automation as a means to cut costs and expedite development, its effectiveness depends on various factors like timing, context, and approach.
To address this, the researchers conducted a multivocal literature review of 78 sources, including formal and grey literature. They categorized factors influencing automation decisions into five groups and identified prevalent ones such as regression testing needs, economic factors, and SUT maturity. The study concludes that current decision-support in software test automation offers reasonable advice for the industry. As a practical outcome, the findings have been synthesized into a checklist for practitioners. However, there is a recommendation to develop systematic, empirically validated decision-support approaches, as existing advice often lacks systemization and is based on weak empirical evidence. Our work differs from this one in terms of goals and research method, but we share with it the procedure of grey literature analysis, which we took inspiration from for our study.

\subsection{Secondary Studies on AI in Test Automation}

In the literature, there are numerous reviews and surveys focusing on Test Automation via AI/ML.
For instance, Trudova et al.~\cite{enase20} conducted a systematic literature review to explore the role of AI/ML in TA. Their findings based on 34 primary studies on AI implementation in software testing reveal that most studies in the literature investigate the application of ML and computer vision techniques to reduce manual intervention in software testing and enhance the effectiveness and reusability of test suites. In particular, the activities which could be improved by the adoption of AI techniques are as follows: test case generation, test oracle generation, test execution, test data generation, test results reporting, test repair, test case selection, flaky test prediction, and test order generation. Also in their case, as in our analysis, the analyzed papers mainly addressed test case generation.

The study conducted by Lima et al.~\cite{9141124} presents a concise overview of the current state of software testing, with a specific focus on the integration of ML and AI algorithms. It evaluates the progress made in AI and ML techniques for software testing over the last three years, drawing from databases including Scopus Elsevier, Web of Science, and Google Scholar. The algorithms are classified into white-box, grey-box, and black-box testing, with an examination of their respective application domains.
The authors conclude that black-box testing emerges as the predominant approach in software testing involving AI. Furthermore, within black-box testing, all three ML methods---supervised, unsupervised, and reinforcement learning---are commonly utilized. Notably, techniques such as clustering, Artificial Neural Networks, and Genetic Algorithms find extensive use in tasks such as fuzzing and regression testing. Differently from this study, which aims to infer the AI and ML algorithms used in software testing, we focused on other aspects as well because grey literature is more descriptive and often lacks technical details.

Our work falls within the scope of these secondary studies. Unlike previous works, we focused on the grey literature to gain insights into practitioners' perceptions of AI/ML adoption in TA. Moreover, we corroborated our taxonomies with interviews with industrial professionals.
\section{Conclusions and Future Work}\label{sec:conclusions}

Artificial intelligence proposes to revolutionize the way we develop and test software systems. Novel tools and testing platforms are being proposed every year, however, to date, little is still known about AI-based test automation, what problems it addresses, what solutions it offers, what tools are available, and for what scope. 

To fill this gap, in this paper, we present a multi-year study of the grey literature concerning AI solutions for test automation. We manually analyzed several thousands of documents from which we retrieved many problems about different aspects of the automated testing process. Moreover, our taxonomy includes the solutions that are used to mitigate such problems and the list of the most popular tools available. 
\changed{The insights provided by our taxonomies were corroborated by five interviews with industrial practitioners, who also provided further insights about the usage of AI in TA.} 

Future research directions consist of conducting a multi-vocal literature review by integrating the findings gathered from the grey literature with those of the white literature.
It would be also interesting to conduct controlled experiments with existing AI-enhanced tools, to quantify the benefits they provide and to validate the observed connections with the TA problems and the investigated solutions.
\section{Acknowledgments}\label{sec:acks}

This research was funded by the Bavarian Ministry of Economic Affairs, Regional Development and Energy.
We extend our deepest gratitude to the professionals who generously took the time to participate in our interviews.


\bibliographystyle{cas-model2-names}

\balance
\bibliography{paper}



\end{document}